\documentclass[twocolumn,showpacs,preprintnumbers,amsmath,amssymb]{revtex4-1}

\usepackage{subfigure}
\usepackage{amsmath}
\usepackage{epsfig,psfrag}
\usepackage{pstricks}
\usepackage{dcolumn}
\usepackage{bm}
\usepackage{graphicx}
\usepackage{color}
\usepackage{version}

\newcommand{\comm}[2]{\left[#1,#2\right]}
\newcommand{\ket}[1]{\left|#1\right>}

\newcommand{\av}[1]{\left<#1\right>}

\begin{document}
\title{Admittance of the SU(2) and SU(4) Anderson quantum RC circuits}
\author{Michele Filippone$^1$}
\author{Karyn Le Hur$^2$}
\author{Christophe Mora$^1$}
\affiliation{$^1$~Laboratoire Pierre Aigrain, \'Ecole Normale
  Sup\'erieure, Universit\'e Paris 7 Diderot, 
CNRS; 24 rue Lhomond, 75005 Paris, France}
\affiliation{$^2$~Centre de Physique Th\' eorique, \' Ecole Polytechnique, CNRS, 91128 Palaiseau C\'edex, France}

\begin{abstract}
We study the Anderson model as a description of the quantum RC circuit for spin-$1/2$ electrons and a single level connected to a single lead. Our analysis relies on the Fermi liquid nature of the ground state which fixes the form of the low energy effective model. The constants of this effective model are extracted from a numerical solution of the Bethe ansatz equations for the Anderson model. They allow us to compute the charge relaxation resistance $R_q$ in different parameter regimes. In the Kondo region, the peak in $R_q$ as a function of the magnetic field is recovered and proven to be in quantitative agreement with previous numerical renormalization group results. In the valence-fluctuation region, the peak in $R_q$ is shown to persist, with a maximum value of $h/2 e^2$, and an analytical expression is obtained using perturbation theory. We extend our analysis to the SU(4) Anderson model where we also derive the existence of a giant peak in the charge relaxation resistance.
\end{abstract}

\pacs{71.10.Ay, 73.63.Kv, 72.15.Qm}

\maketitle

\section{Introduction}
\begin{figure}[b]
  \includegraphics[width = \linewidth]{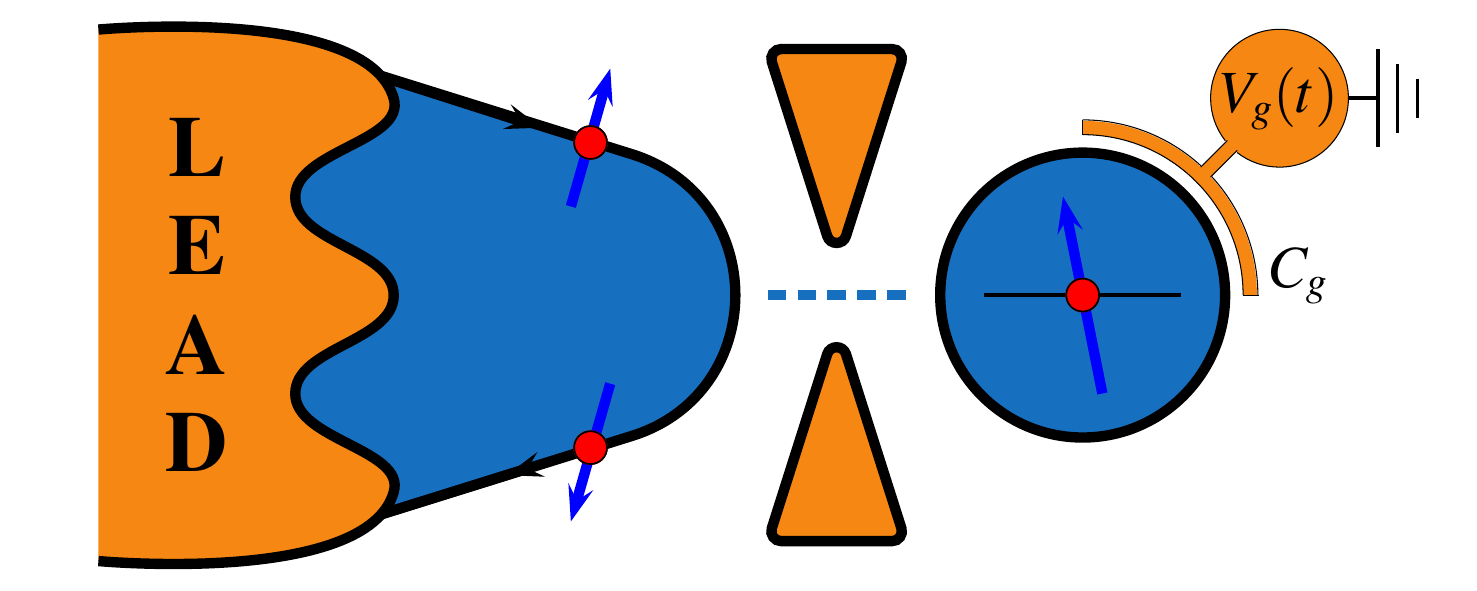}
  \caption{Schematic view of the Anderson quantum RC circuit. Spin-degenerate electrons tunnel, through a quantum point contact, between a reservoir lead and a single level with a local interaction. An oscillating voltage $V_g(t)$ is applied to the dot via a metallic gate of geometric capacitance $C_g$.}\label{fig:rc}
 \end{figure}

 High frequency transport experiments aim to control and probe the coherent motion of electrons in real time \cite{deblock2003detection,petta2005coherent,koppens2006driven,bocquillon2013}. Continuous technological progress have paved the way for the study of dynamical properties of mesoscopic systems. A typical example is the quantum RC circuit \cite{gabelli2006,gabelli2012coherent,buttiker1993b}, where a quantum dot is connected to the quantum Hall edge states of a two-dimensional electron gas (2DEG) through a quantum point contact (QPC). As illustrated in Fig.~\ref{fig:rc}, the quantum dot forms a mesoscopic capacitor with a top metallic gate and the charge in the dot can be changed periodically by applying an AC drive through the gate voltage $V_g$. For large gate voltage modulations, this device acts as a single electron emitter \cite{feve2007,mahe2010,parmentier2011,grenier2011single,bocquillon2012} and the effect of sweeping the last occupied level of the dot across the Fermi energy has received a broad theoretical attention~\cite{moskalets2008,keeling2008,mathias2011,battista2012}. The quantum RC circuit is also promising for efficient charge readout in a quantum dot device~\cite{nigg2009} or to detect topological excitations \cite{garate2012,golub2012charge}. The electron dynamics in the presence of interactions in the dot \cite{brouwer2005,nigg2006} and its spin/charge separation \cite{splett2010} have been also studied. For small metallic islands the problem has been addressed at intermediate temperatures  \cite{rodionov2009} and in the many channel case~\cite{etzioni2011}. In particular, the two-channel case has been argued to exhibit non-Fermi liquid behavior \cite{mora2012low,dutt2013}. Novel perspectives have been opened by recent experiments \cite{delbecq2011,frey2012,petersson2012,schroer2012} where a significant dipole coupling between a microwave superconducting resonator and a quantum dot has been demonstrated.

The low frequency admittance for the current $I$ from the dot to the lead can be matched with the corresponding formula for a classical RC circuit
\begin{equation}\label{admi}
\frac{I(\omega)}{V_g (\omega)} = -i \omega C_0 (
1+ i \omega \, C_0 R_q ) + {\rm O} (\omega^3).
\end{equation}
This allows one to define a \emph{quantum capacitance} $C_0$ and 
a \emph{charge relaxation resistance} $R_q$ for the AC admittance of the system. This formula is related to the dynamic charge susceptibility of the dot $\chi_c(\omega)$ by the relation $I(\omega)/V_g(\omega)=-i\omega e^2\chi_c(\omega)$. $\chi_c(t-t')=i/\hbar\theta(t-t')\av{\comm{n(t)}{n(t')}}$ is the linear response function of the total occupancy $n$ of the dot to a change in the gate voltage. Identifying term by term the low frequency expansion of $\chi_c(\omega)$ with Eq. \eqref{admi}, the definitions of the quantum capacitance and the charge relaxation resistance are obtained
\begin{align}\label{def}
C_0&=e^2\chi_c(0)\,,&R_q&=\left.\frac{e^2\mbox{Im}\chi_c(\omega)}{\omega C_0^2}\right|_{\omega\rightarrow0}\,.
\end{align}

These quantities have raised a large interest from the theoretical point of view starting with the seminal works of B\"uttiker, Pr\^etre and Thomas \cite{buttiker1993,buttiker1993b,pretre1996dynamic}. In the quantum regime, the quantum capacitance $C_0$ provides information on the level structure~\cite{cottet2011} of the quantum dot. For a single channel in the QPC connecting the dot and the lead, $R_q$ is universally fixed to $h/2e^2$ regardless of the QPC transparency. This prediction has been experimentally demonstrated~\cite{gabelli2006}. It coincides with the lead-reservoir interface resistance~\cite{buttiker1986} relevant in DC transport~\cite{nigg2008quantum,buttiker2009role}.

The universality of $R_q$ still holds if interactions in the dot \cite{mora2010} or not too strong interactions in the lead \cite{hamamoto2010,*hamamoto2011quantum} are taken into account in an exact manner. Increasing the size of the dot results in a mesoscopic crossover for $R_q$ from $h/2e^2$ to $h/e^2$~\cite{mora2010}. For strong enough interactions in the lead, {\it i.e.} a Luttinger parameter below $1/2$, the system undergoes a Kosterlitz-Thouless phase transition to an incoherent regime where $R_q$ is no longer quantized~\cite{hamamoto2010,*hamamoto2011quantum}.

In this paper, we investigate the AC linear regime of the quantum RC circuit where electrons carry a spin degree of freedom, as represented in Fig.~\ref{fig:rc}, and the system is described by the Anderson model. Throughout the paper, we shall focus on the regime where the local interaction term $U$ is much larger than the hybridization energy $\Gamma$ such that charge fluctuations are small except at the charge degeneracy points, {{\it i.e.} the Coulomb peaks. It includes in particular the Kondo regime where the spin on the dot is strongly correlated with the Fermi sea in the reservoir lead. Our analysis shall also include the more exotic SU(4) Kondo regimes relevant for dots with an additional orbital degree of freedom \cite{borda2003,*lehur2003,*zarand2003kondo,*lopez2005,*choi2005,jarillo2005orbital,*makarovski2007,*tettamanzi2012}. 

The charge relaxation resistance of the Anderson model has been recently investigated by numerical renormalization group (NRG) calculations~\cite{lee2011}, where it was shown that $R_q$ develops a giant peak at zero temperature for Zeeman energies of the order of the Kondo temperature $T_K$. An analytical description of this peak has been given in the Kondo scaling limit~\cite{filippone2011}, on the basis of a Fermi liquid description valid at low temperature, in quantitative agreement with the NRG results. It also predicts the disappearance of the resistance peak at the particle-hole symmetric point, $\varepsilon_d = -U/2$, where $\varepsilon_d$ denotes the single-orbital energy on the dot. The Fermi liquid approach~\cite{filippone2012} is based on the identification of the low energy effective model, consistent with the Friedel sum rule. It allows one to derive a generalized Korringa-Shiba relation~\cite{shiba1975}
\begin{equation}\label{korringa}
\lim_{\omega\rightarrow0}\frac{\mbox{Im}\chi_c(\omega)}\omega=\hbar\pi\left(\chi_{\uparrow}^2+\chi_{\downarrow}^2\right),
\end{equation}
which relates the dynamical charge susceptibility $\chi_c(\omega)$ to the static ones $\chi_{\sigma}=-\partial \av{n_\sigma}/\partial \epsilon_d$. $\av{n_\sigma}$ denotes the static occupancy of the dot for spin $\sigma$. A similar relation was previously obtained for the spin susceptibility using the same Fermi liquid arguments~\cite{garst2005}. Comparing Eq.~\eqref{korringa} with Eq.~\eqref{def}, a general formula for the charge relaxation resistance
\begin{equation}\label{rqintro}
R_q=\frac{h}{4e^2}\left(1+\frac{\chi_m^2}{\chi_c^2}\right)
\end{equation}
is extracted where we have introduced $\chi_c=\chi_{\uparrow}+\chi_{\downarrow}$, the total charge susceptibility and $\chi_m=\chi_{\uparrow}-\chi_{\downarrow}$, the \emph{charge-magneto susceptibility}~\cite{filippone2011}. $\chi_m$ should be clearly distinguished from the spin susceptibility, which is the derivative of the magnetization with respect to the magnetic field. The whole point of Eq.~\eqref{rqintro} is that $R_q$, a dynamical quantity, is expressed in terms of static quantities computable by Bethe ansatz (BA). Deviations from universality occur in Eq.~\eqref{rqintro} when $\chi_m \ne 0$, that is when both the particle-hole and the SU(2) spin symmetries are broken.
\begin{figure}
 \begin{center}
    \subfigure[]{\includegraphics[width = \linewidth]{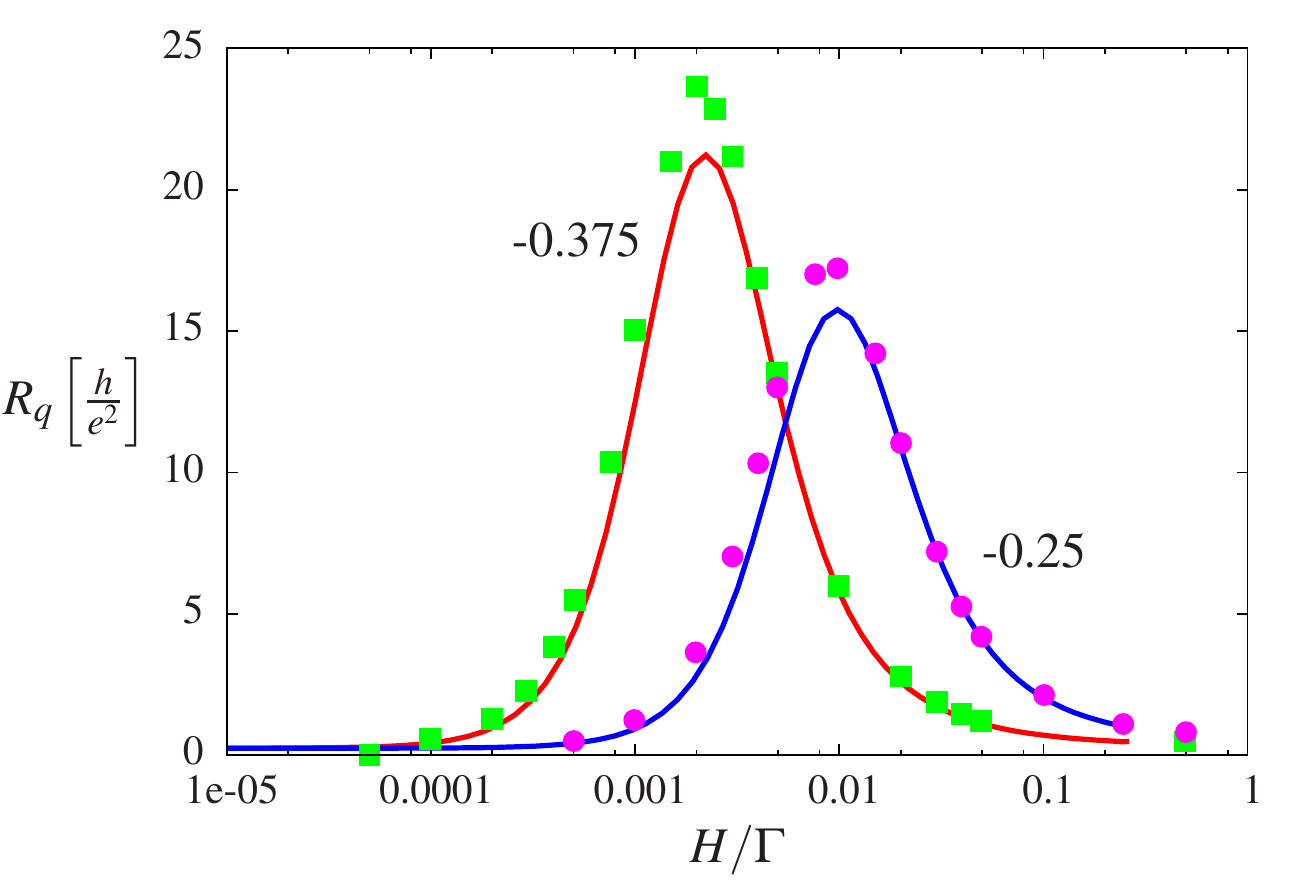}}
    \subfigure[]{\includegraphics[width = \linewidth]{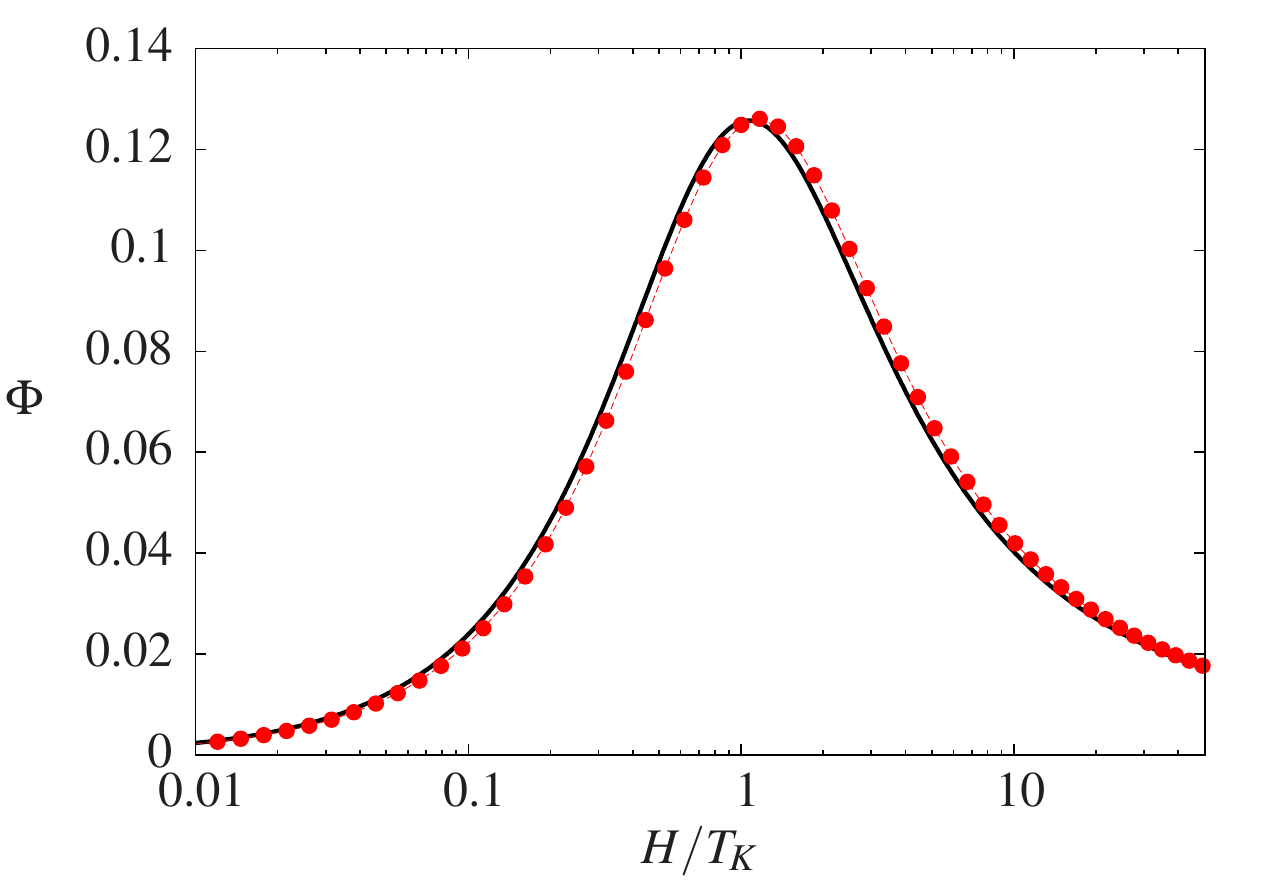}}    
    \caption{a)  Comparison of $R_q$ as a function of the magnetic field between NRG calculations (dots) (extracted from Ref.~\cite{lee2011}) and BA results (solid lines) for different $\varepsilon_d/U$ and $U/\Gamma=20$, showing good agreement. b) Comparison of the universal function $\Phi_0(H/T_K)$ (solid) derived in the Kondo scaling limit in Ref. \cite{filippone2011} with the function $\Phi$ (dots) derived by Bethe ansatz calculations from Eq. (\ref{fphi}) for $U/\Gamma=20$ and close to the mixed-valence region ($\varepsilon_d/U=-0.2$).}\label{fig:complee}
   \end{center}
  \end{figure}

In this work, we extend the Fermi liquid analysis to different parametric regimes by solving numerically the BA equations for the ground state~\cite{wiegmann1983,tsvelick1983,kawakami1982ground} and computing the static susceptibilities $\chi_c$ and $\chi_m$ appearing in Eq.~\eqref{rqintro}. In the Kondo region, the robustness of the scaling form proposed in Ref.~\cite{filippone2011} is tested for finite parameters of the Anderson model, as shown in Figs.~\ref{fig:complee}.b and~\ref{fig:F}. We confirm notably in Fig.~\ref{fig:complee}.a that the Fermi liquid result Eq.~\eqref{rqintro} agrees nicely with the NRG calculations of Ref.~\cite{lee2011}. Out of the Kondo regime, departures from universality of the charge relaxation resistance as a function of the magnetic field were also shown within the Hartree-Fock approximation \cite{nigg2006}. Extending the BA calculations to the mixed-valence, empty orbital and valence-fluctuation regimes, we find that the peak in the charge relaxation resistance survives in these regimes, although its magnitude decreases in size with $\varepsilon_d/U$. Interestingly, even far in the valence-fluctuation region, {\it i.e.} for large $\varepsilon_d/U$ and $H \simeq \varepsilon_d$, the peak is still present: $R_q$ varies between $h/4e^2$ and $h/2e^2$ as a function of the magnetic field. The corresponding universal function for $R_q$, represented in Fig.~\ref{fig:scaling}, is derived analytically using perturbation theory and shown to agree with the BA calculations. In this region, the peak in $R_q$ is not generated by breaking the Kondo singlet, but by the transition between different charge states of the dot.

\begin{figure}[t]
\begin{center}
\includegraphics[width=\linewidth]{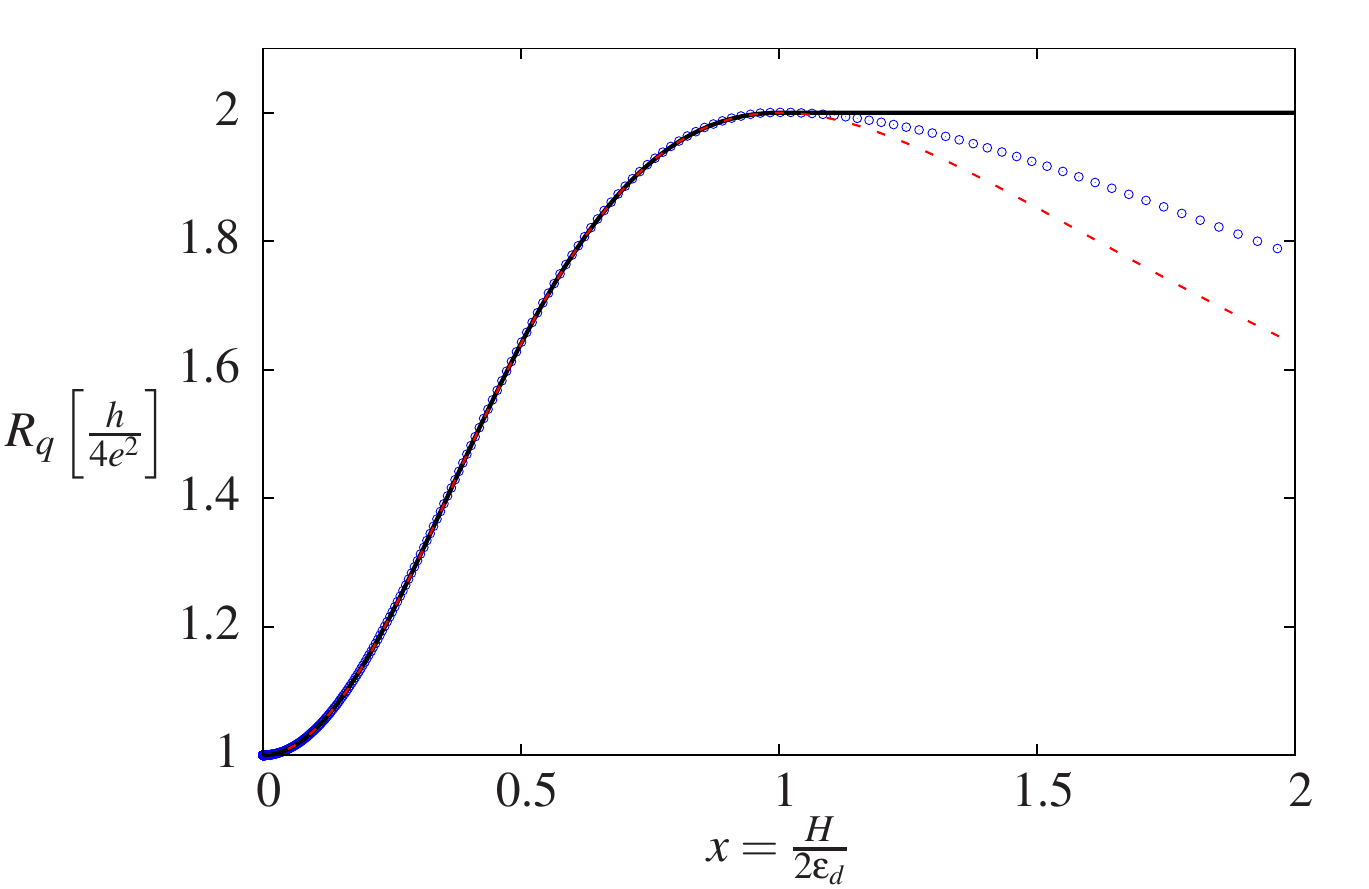}
\caption{Scaling form of the charge relaxation resistance $R_q$ in the limits $\Gamma\ll \varepsilon_d\ll U$ (solid line) and $\Gamma\ll U\ll\varepsilon_d$ (dashed line). In circles, the resistance $R_q$ is plotted for $U=\varepsilon_d$.}\label{fig:scaling}
\end{center}
\end{figure}

We finally give a further application of the Fermi liquid approach~\cite{filippone2011,filippone2012} by considering an additional orbital degeneracy in the dot responsible for SU(4) Kondo behavior at low energy~\cite{choi2005,jarillo2005orbital,*makarovski2007,*tettamanzi2012}. The existence of a Fermi liquid ground state~\cite{affleck1990current,bazhanov2003} in the case of a SU(4) symmetry allows us to derive an analog of Eq. (\ref{rqintro}). In the Kondo scaling limit, we predict, similarly to the SU(2) case, a giant peak in the charge relaxation resistance. 

The paper is organized as follows. Sec. \ref{sec:fermi} explains how $C_0$ and $R_q$ are calculated by solving the BA equations for the ground state of the Anderson model once the Fermi liquid fixed point is determined. In Sec. \ref{sec:testscaling} we study the range of validity of the Kondo scaling limit obtained in \cite{filippone2011}. In Sec. \ref{sec:mixed}, we analyze the new scaling forms of $R_q$ in the valence-fluctuation region. The peak in the charge relaxation resistance for a SU(4) symmetric Anderson model in the Kondo limit is presented in Sec. \ref{sec:carbon}.


\section{Fermi liquid picture}\label{sec:fermi}
The relevant model to describe the quantum RC circuit in Fig. \ref{fig:rc}, when the dot level spacing is sufficiently large and the transport is not spin-polarized, is the Anderson model \cite{lee2011,filippone2011} 
\begin{equation}\label{am}
\begin{split}
H_{\rm AM} &= \sum_{\sigma,k} \varepsilon_{k\sigma} 
 c^\dagger_{k\sigma} c_{k\sigma}
+  \sum_\sigma\varepsilon_{d\sigma} 
\, { n_\sigma} \\
& + U {n}_{\uparrow} {n}_{\downarrow}
+ t \sum_{k,\sigma} \left( c_{k\sigma}^\dagger  d_\sigma +  d_\sigma^\dagger
c_{k\sigma} \right).
\end{split}
\end{equation}
This Hamiltonian describes a single level, whose double occupation costs a charging energy
$U$, weakly coupled to a non-interacting electron bath. 
The operators $c_{k\sigma}$ and $d_\sigma$ annihilate electrons of spin
 $\sigma$ on the lead and on the dot respectively.
The lead electrons are characterized by the single-particle 
dispersion relation $\varepsilon_k$ with a constant density of states $\nu_0$.
 The total electron occupancy 
of the dot is $n = n_{\uparrow} + n_{\downarrow}$
with $n_\sigma =  d_\sigma^\dagger d_\sigma$.
The geometric capacitance $C_g$ and the tunable electrostatic coupling $V_g$ between the dot and the metallic top gate enter in Eq. (\ref{am}) through the interaction, or charging, energy $U=e^2/C_g$
 and the single-electron orbital energies $\varepsilon_{d\sigma} = - e V_g-\sigma H/2$
where  $H$ is the external magnetic field. $t$ is the amplitude for electron tunneling between the dot and the lead and we assume the hybridization constant $\Gamma = \pi \nu_0 t^2$ to be independent of the magnetic field
\footnote{For QPCs in 2DEGs, this approximation is justified as the tunneling becomes sensitive to the magnetic field when the Zeeman splitting of the transverse modes in the QPC becomes of the order of their level spacing, which is higher than $T_K$ \cite{lehur2001}. The regime studied in Section \ref{sec:mixed} is then relevant for setups where the tunneling is weakly dependent on the magnetic field, as for carbon nanotube dots. We neglect this dependence here, but it must be kept in mind that it controls the closing of the anti parallel spin channel bringing $R_q$ back to $h/2e^2$ in the large magnetic field limit.}.

\begin{figure}[t!!]
\begin{center}
\subfigure[]{\includegraphics[width= 7.5cm]{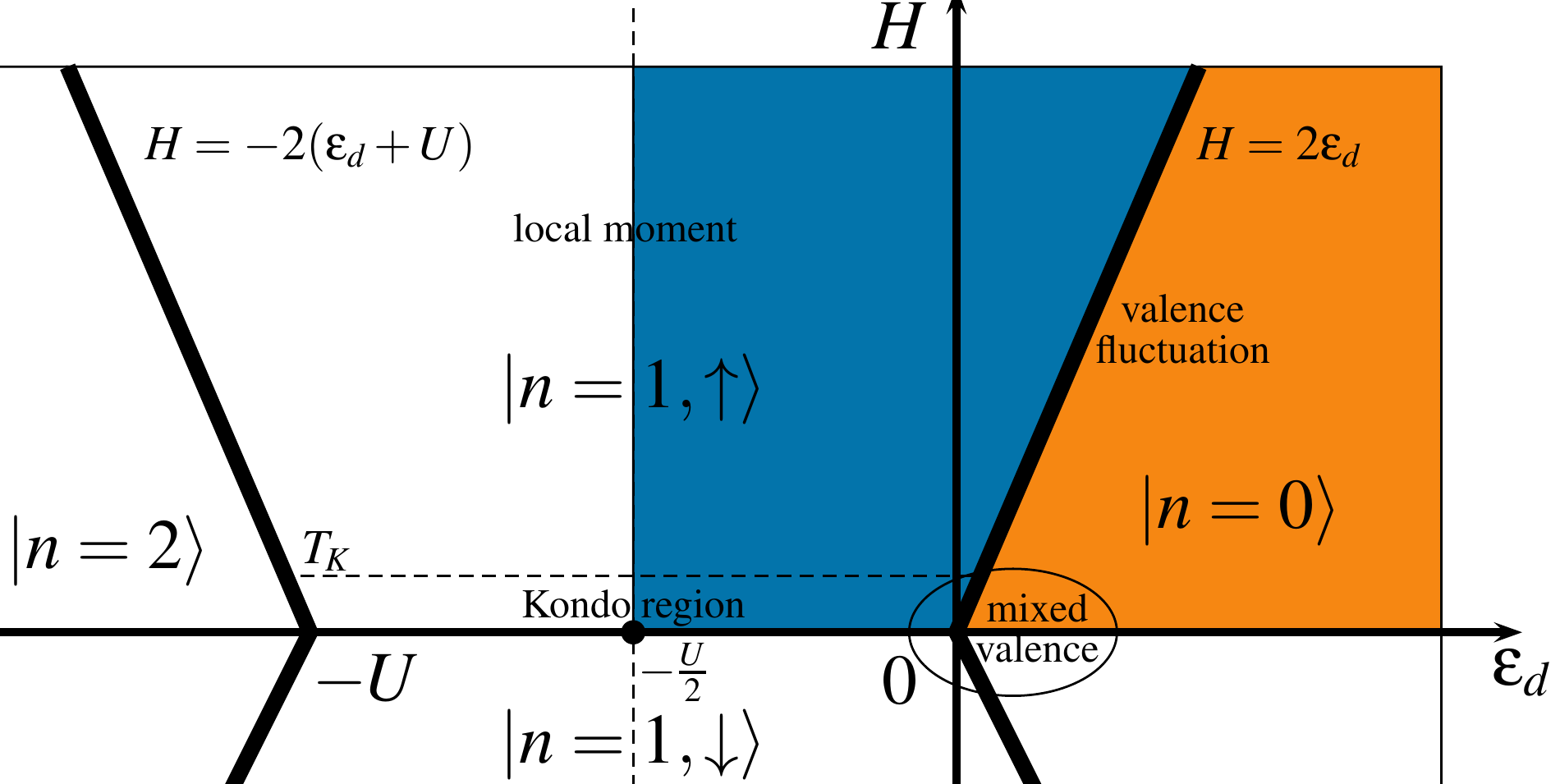}}
\subfigure[]{\includegraphics[width=7.5cm]{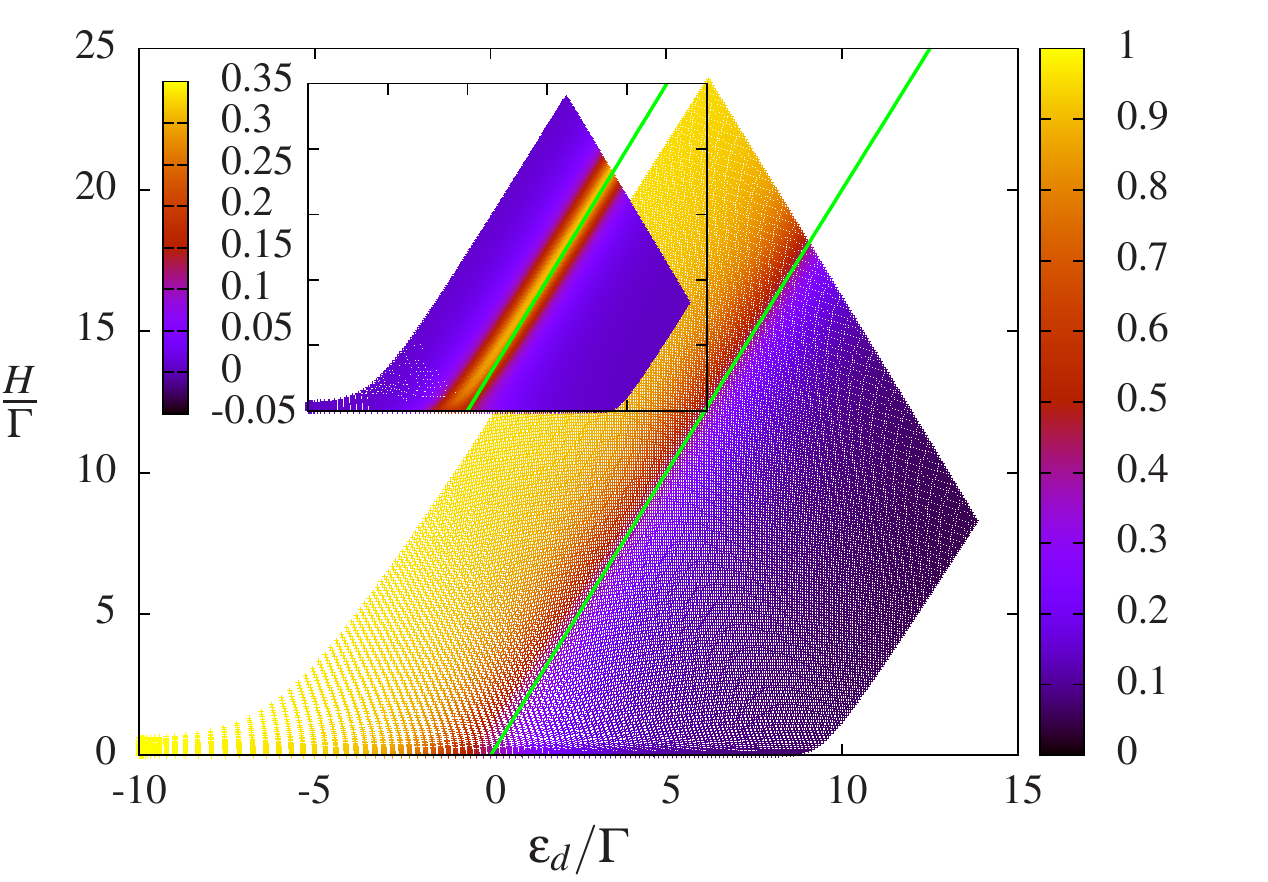}}
\subfigure[]{\includegraphics[width=7.5cm]{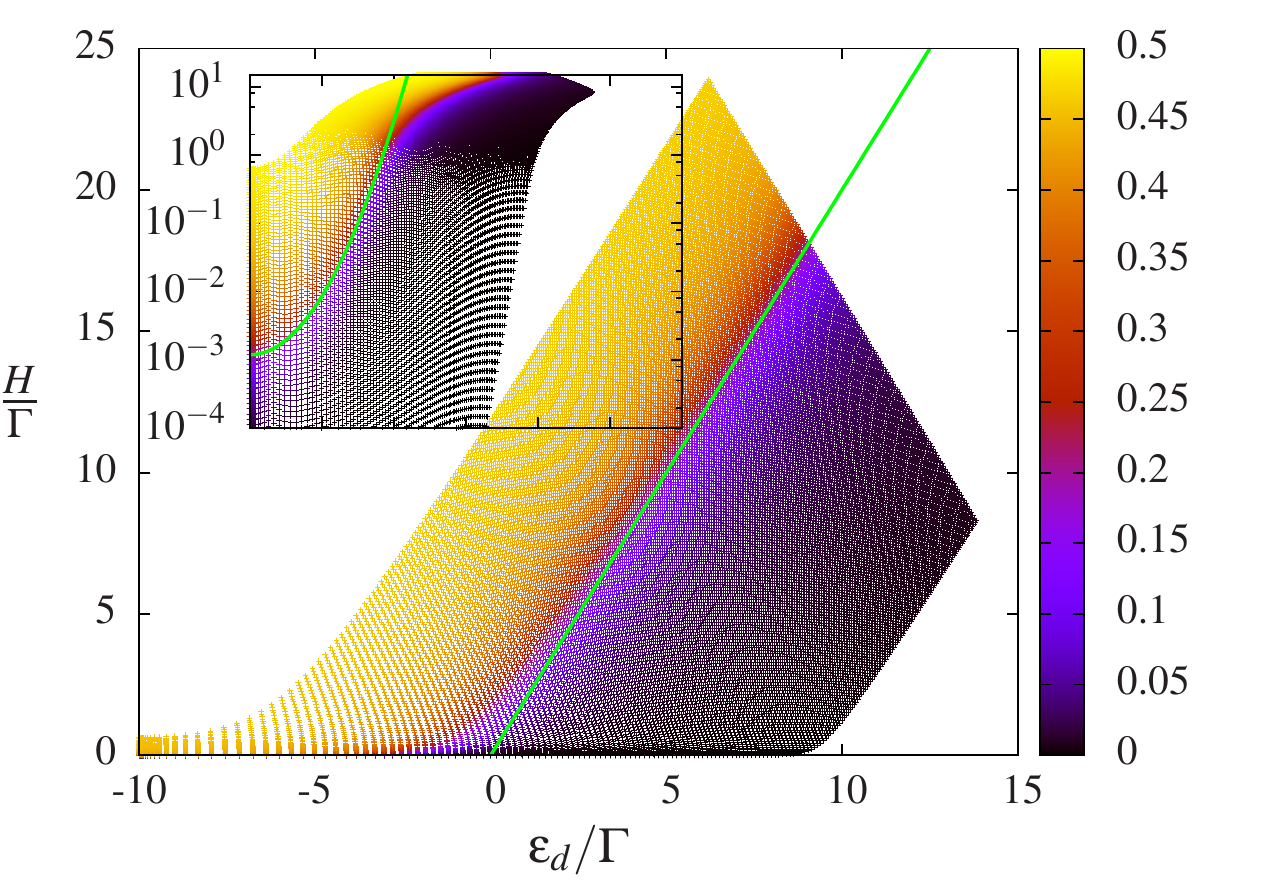}}
\caption{a) Phase diagram of the isolated dot in the presence of a magnetic field. b) and c) are computed from the Bethe ansatz equations summarized in Appendix~\ref{app:BA}. The domain $\varepsilon_d < -U/2$ is deduced from the colored domain ($\varepsilon_d < -U/2$) by particle-hole symmetry. b) Occupation number of the dot $\av{n}$, for $U/\Gamma=20$, reproducing the phase diagram in a). The boundaries between charge states are smooth functions. The smaller panel represents the static charge susceptibility $\chi_c$ with its Coulomb peaks. c) Magnetization $\av{m}$ of the dot, resembling the phase diagram in a), except at low energy. The difference is more visible in the smaller panel (logarithmic scale), where the local moment is screened below the Kondo temperature $T_K$ (solid line).}\label{fig:phase}
\end{center}
\end{figure}
 
It is a well established fact that the Anderson model behaves as a Fermi liquid at zero temperature \cite{haldane1978scaling,krishna1980} for all values of the single-electron
orbital energies $\varepsilon_{d\sigma}$. Moreover, the phase shift of quasi-particles at the Fermi
energy is fixed by the dot occupancy through the Friedel sum rule~\cite{langreth1966}
\begin{equation}\label{friedel}
\av{n_\sigma}=\frac{\delta_\sigma}\pi.
\end{equation}
For a time-dependent gate voltage $\varepsilon_d (t)=
\varepsilon_d^0+\varepsilon_\omega\cos(\omega t)$, 
the form of the Hamiltonian follows from a quasi-static 
approximation~\cite{filippone2011,filippone2012}, consistent with the Friedel sum rule
Eq.~\eqref{friedel}
\begin{equation}\label{fermiliquid}
  H_{FL}=\sum_{k\sigma}\varepsilon_ka^\dagger_{k\sigma}a_{k\sigma}+\epsilon_\omega\cos\omega t\sum_\sigma\frac{\chi_{\sigma}}{\nu_0}\sum_{k,k'}a^\dagger_{k\sigma}a_{k'\sigma},
\end{equation}
where the $a_{k\sigma}$ operators describe quasiparticle states with a 
phase shift $\delta_\sigma(\varepsilon_d^0)$ with respect to the original fermions
$c_{k\sigma}$.

The dot variables have disappeared from the effective Hamiltonian Eq.~\eqref{fermiliquid},
although the memory of the dot is kept in the static charge susceptibilities 
$\chi_{\sigma}$. Practically, the occupation number  $\av n=\av{n_\uparrow}+\av{n_\downarrow}$
and the magnetization $\av m=(\av{n_\uparrow}-\av{n_\downarrow})/2$ are static observables
and they are  obtained by solving numerically the BA equations summarized in Appendix~\ref{app:BA}.

The study of $\av n$ and $\av m$ identifies four regimes shown in Fig. \ref{fig:phase}a. A finite hybridization $\Gamma$ between the dot and the lead smoothens the boundary lines between the different charge and spin states of the dot, as seen in Figs.~\ref{fig:phase}b and \ref{fig:phase}c.  The region where the charge is equal to 1 and the magnetization to 1/2 is called the \emph{local-moment} region. The transition to the empty (or doubly occupied) orbital regimes, where the charge is held fixed to zero (or two), takes place in the \emph{valence-fluctuation} region. The \emph{valence-fluctuation} region is signaled by a Coulomb peak in the charge susceptibility $\chi_c$ (visible in the smaller panel of Fig. \ref{fig:phase}.b) which defines the frontiers between the different Coulomb-blocked regions with zero, one or two charges. For Zeeman energies below $H_1=\Gamma U/(U+2\varepsilon_d)$, the \emph{mixed-valence} region is entered and the Coulomb peak deviates from the $H=2\varepsilon_d$ line touching the $H=0$ axis at $\varepsilon_d^*=0$, where $\varepsilon_d^*=\varepsilon_d+\Gamma/\pi\ln(\pi e U/4\Gamma)$ is the renormalized orbital energy of the dot~\cite{haldane1978scaling,wiegmann1983}. This deviation is presented in Fig.~\ref{fig:chic}. The magnetization, shown in Fig. \ref{fig:phase}.c, shows a different behavior from the charge occupation of the dot. The transition line between a magnetized and a non-magnetized state penetrates in the local-moment region following the Kondo temperature \cite{tsvelick1983}
\begin{equation}\label{tk}
T_K=2\sqrt{\frac{U\Gamma}{\pi e}}e^{\frac{\pi\epsilon_d(\epsilon_d+U)}{2U\Gamma}}.
\end{equation}
This is the signature of a strongly correlated ground state where the lead electrons screen the spin of the dot by forming a many-body Kondo singlet~\cite{hewson1997kondo}. In general, this state cannot be described by standard perturbation techniques. In this paper, we circumvent this difficulty by solving the BA equations for the static quantities, combined with a Fermi liquid approach to access the low frequency behavior of the dynamical charge susceptibility $\chi_c(\omega)$.


\subsection{The quantum capacitance $C_0$}\label{sec:capa}

The quantum capacitance $C_0=e^2\chi_c$ appears to leading
order in the frequency expansion of Eq.~\eqref{admi}. 
The static charge susceptibility $\chi_c$ can be
calculated from the Bethe ansatz solution and
is plotted in the inset of Fig. ~\ref{fig:phase}.b.
It exhibits strong Coulomb peaks at charge degeneracy
points for $U \gg \Gamma$, as a result of charge
quantization.
 $\chi_c$ is also represented in Fig.~\ref{fig:chic} 
as a function
of the gate voltage for different values of
the magnetic field.  Fig.~\ref{fig:chic} illustrates 
in particular
that $\chi_c$ is insensitive to the magnetic field until
the Zeeman energy is of the order of $\Gamma$.
In the Kondo region, the Kondo temperature $T_K$ is much
smaller than $\Gamma$, and the peak in the charge relaxation
resistance  $R_q$ thus develops in a region where the static 
charge susceptibility is independent of the magnetic field.

When the Zeeman energy is above the hybridization constant $\Gamma$, the Coulomb peak starts moving following the $H=2\varepsilon_d$ transition line obtained for the isolated impurity diagram in Fig.~\ref{fig:phase}.a. In this regime, the Coulomb peak has a Lorentzian shape which can be derived analytically by just neglecting the spin down component. This procedure will be presented in Sec.~\ref{sec:mixed}.

We stress that, in contrast with the non-interacting case,
the quantum capacitance is not proportional to the local 
density of states as it is sensitive only to charge
excitations and not to spin excitations. Hence, the Kondo
peak in the density of states, which arises due to spin-flip processes,
has no effect on the quantum capacitance $C_0$.

\begin{figure}[t!]
\begin{center}
    \includegraphics[width = \linewidth]{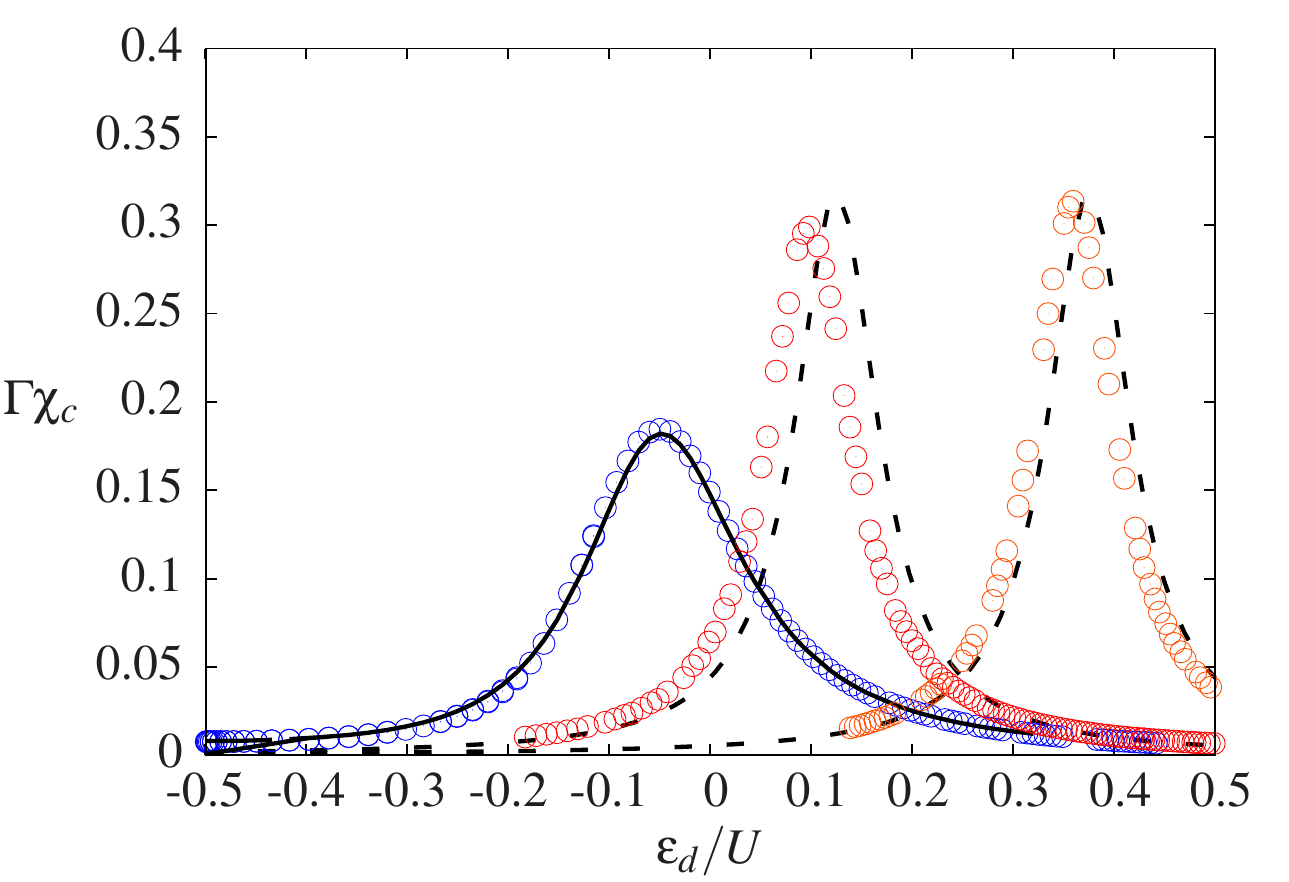}
\caption{Charge susceptibility $\chi_c$ for $U/\Gamma=20$. The circles (left to right) correspond to $H/\Gamma=0.1,~5$ and 15 respectively and show the displacement of the Coulomb peak, also shown in Fig.~\ref{fig:phase}.a. The solid line is obtained for $H/\Gamma=0.0001$ and almost coincides with $H/\Gamma=0.1$, showing the very weak dependence of $\chi_c$ with the magnetic field in the Kondo regime. For higher magnetic fields, $H/\Gamma=5,~15$, $\chi_c$ converges to the Lorentzian form Eq.~\eqref{lorentzian} (dashed lines) derived in the valence-fluctuation region.}\label{fig:chic}
\end{center}
\end{figure}


\subsection{The charge relaxation resistance $R_q$}

The second term in the low-frequency expansion of Eq.~\eqref{admi} 
describes the leading deviation from adiabaticity and introduces the response
time scale RC to a slow drive of the gate voltage. Eq.~\eqref{korringa} derived
in Ref.~\cite{filippone2011,filippone2012} gives the charge relaxation 
resistance $R_q$ for all gates
voltages and magnetic fields. $\chi_c$ and $\chi_m$ are both computed by
solving the Bethe ansatz equation summarized in the appendix.
Before discussing the results for $R_q$ in the different regimes of
parameters, let us note that the particle-hole symmetry of the
Anderson model implies that $\chi_m$ is an odd function of 
$\varepsilon_d+U/2$ and thus vanishes for $\varepsilon_d=-U/2$ . 
As a result, the quantized value $R_q=h/4 e^2$ is obtained at the 
particle-hole symmetric point irrespective of the magnetic field.

In the Kondo region, $R_q$ assumes the form
\begin{equation}\label{rqscaling}
R_q=\frac h{4e^2}\left[1+\left(\frac U\Gamma\right)^4F_0(y)\Phi_0\left(\frac H{T_K}\right)^2\right],
\end{equation}
in the scaling limit $U \gg \Gamma$, $\varepsilon_d^* \gg \Gamma$, 
$H \ll \Gamma$.
The function  $\Phi_0(x)=xf'(x)$, plotted in Fig.\ref{fig:complee}.b, 
is obtained from the universal form of the magnetization  
$m = f(H/T_K)$ for the Kondo model~\cite{andrei1983} with the asymptotic behaviors: 
\begin{equation}\label{asymp}
\begin{split}
\Phi_0 (x) &= \frac{x}{\sqrt{2 \pi {\rm e}}}, \qquad \qquad  x \ll 1, \\
\Phi_0 (x) &= \frac{1}{4} \, \frac{1}{(\ln x)^2}, \qquad \, \, \, \,  x \gg 1,
\end{split}
\end{equation}
where ${\rm e}$ is Euler's number.
The function $\Phi_0$ develops a peak when the magnetic field $H$
is on the order of the Kondo temperature $T_K$. 
The envelope function 
\begin{equation}\label{eqFleading} 
F_0 (y)=  \left( \frac{\pi^2}{8} \right)^2
\,  \frac{y^2 \left(y^2-1\right)^4}{(1+y^2)^2} 
\end{equation}
depends on the asymmetry parameter 
$ y=1+2 \varepsilon_d/U$
and  is shown in Fig.~\ref{fig:F}. It is obtained from the leading order charge
susceptibility (insensitive to the magnetic field for $H \ll  \Gamma$)
and from the derivative of the Kondo temperature $\partial_{\varepsilon_d}
\ln T_K = \pi y/\Gamma$.  $\partial_{\varepsilon_d} \ln T_K$ is an odd function of $y$
such that $F_0(y)$ vanishes at the particle-hole 
symmetric point $y=0$ in agreement with the above discussion.

The robustness of the scaling form Eq.~\eqref{rqscaling} for finite values of
the different parameters of the Anderson model is discussed in the following
Section.


\begin{figure}[t!]
\begin{center}
    \subfigure{\includegraphics[width=8.1cm]{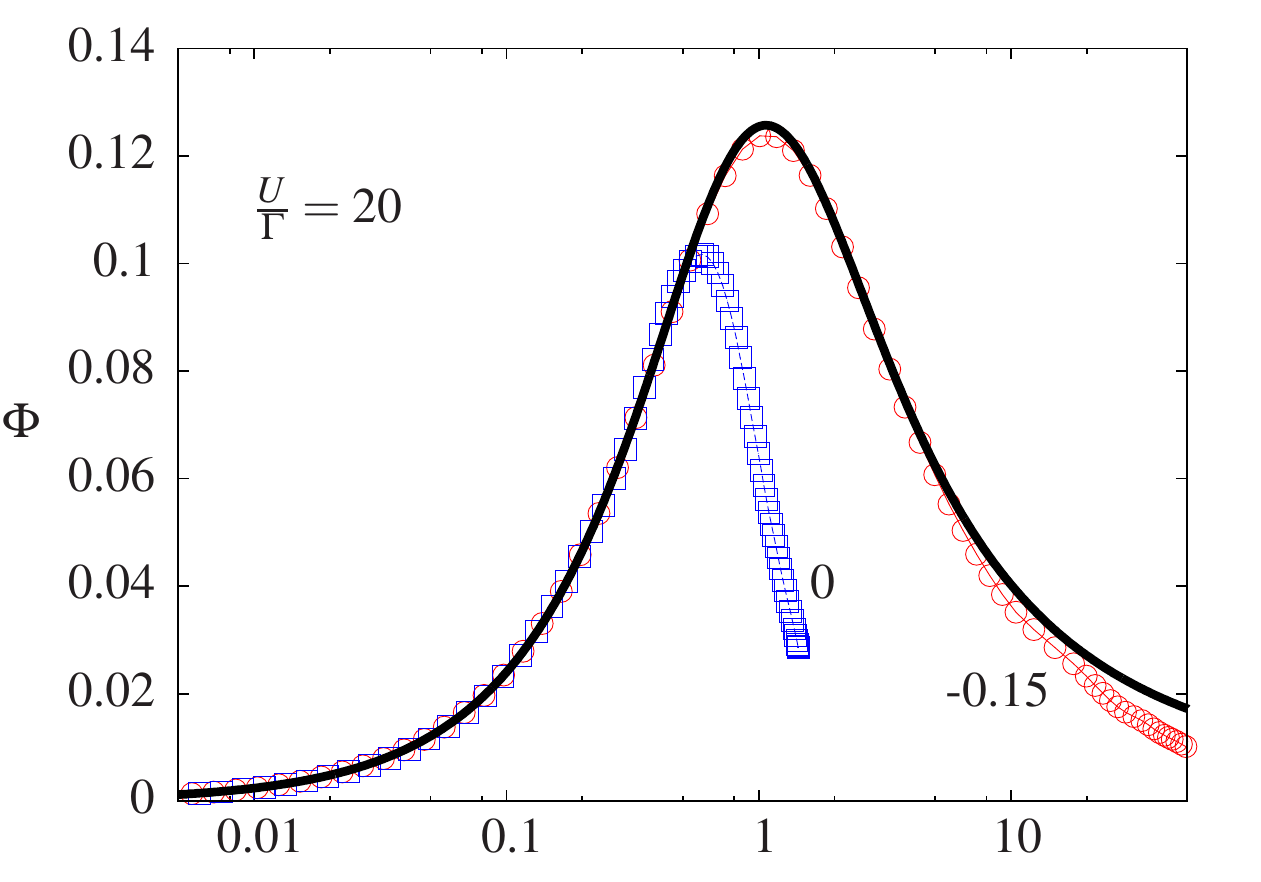}}
    \subfigure{\includegraphics[width=8.1cm]{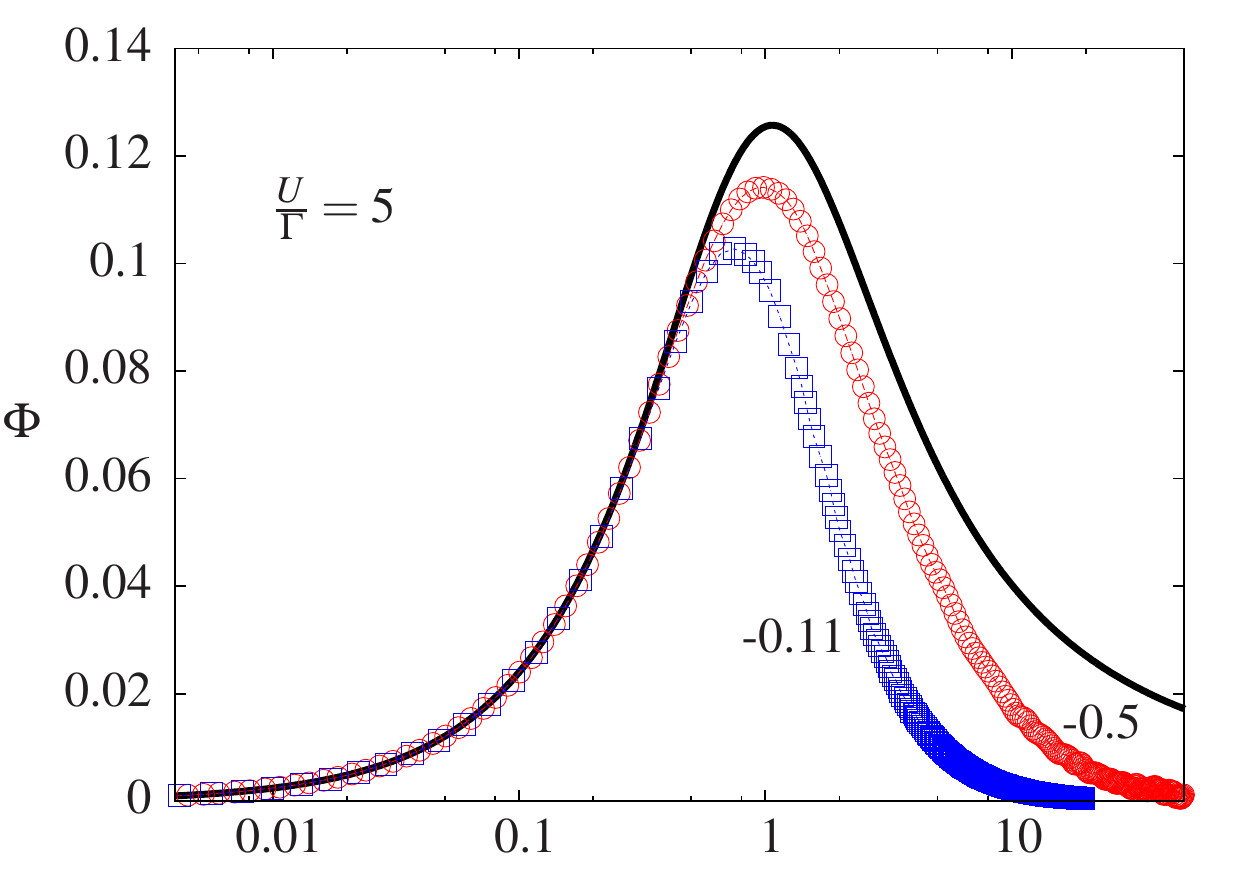}}
    \subfigure{\includegraphics[width=8.3cm]{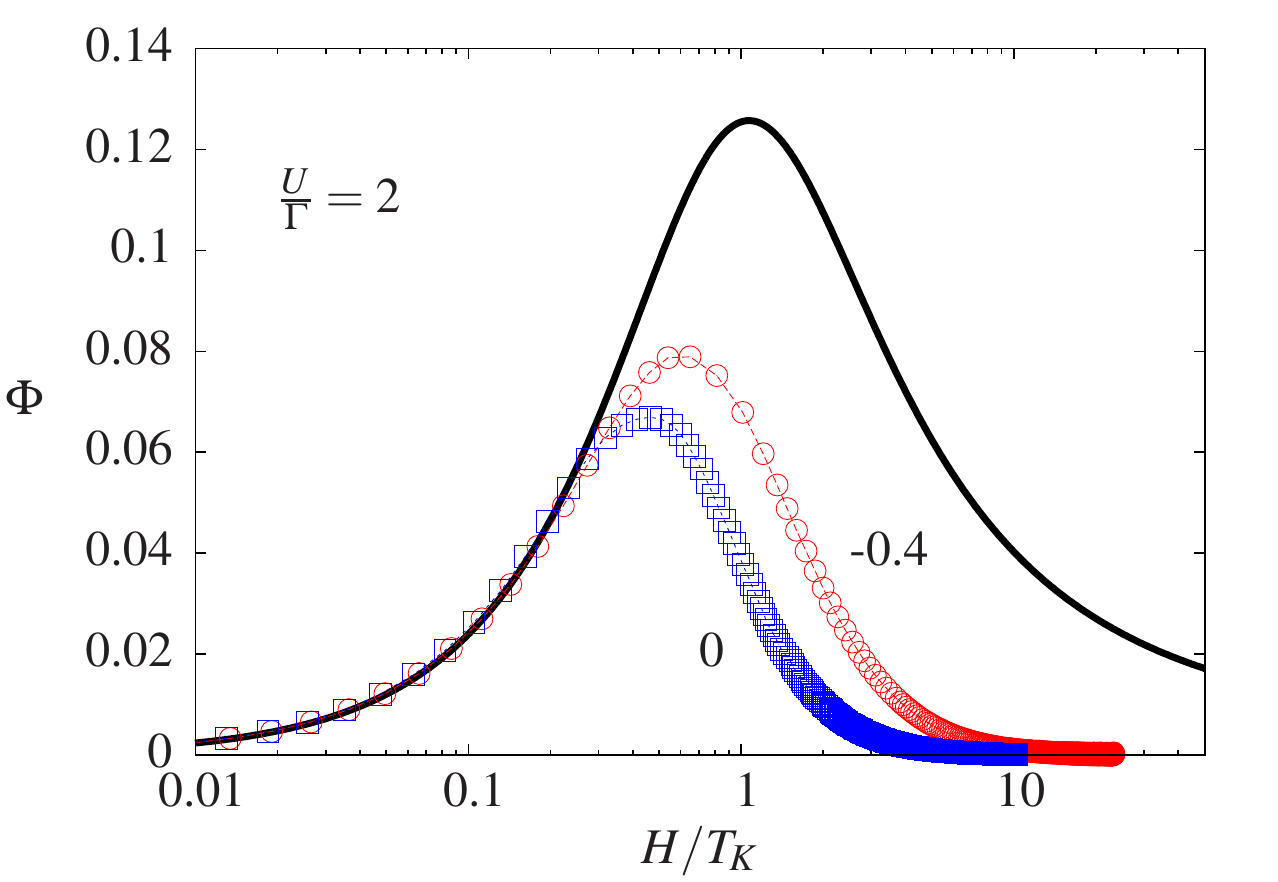}}
\caption{Function $\Phi$ Eq. (\ref{fphi}) for different $\varepsilon_d/U$ (squares and circles) and $U/\Gamma$ calculated by the numerical solution of the Bethe ansatz compared to the universal scaling $\Phi_0$ (solid line). The values of $T_K$ for the numerical data are fixed by matching the values of $\Phi$ at low fields to the linear behavior of $\Phi_0$ Eq. (\ref{asymp}). These are plotted in Fig. \ref{fig:tk}.}\label{fig:phi}
\end{center}
\end{figure}

\section{Scaling form of the charge relaxation resistance}\label{sec:testscaling}

By definition, the scaling form Eq.~\eqref{rqscaling} is only an
asymptotic behavior and it
is of interest to evaluate how quantitative it
is for real systems. In the general case, we extend the definitions
of the two functions 
\begin{align}\label{fphi}
 F&=\left(\frac \Gamma U\right)^4\left(\frac{y\pi}{\Gamma\chi_c}\right)^2,  &   \Phi&=\frac{\Gamma\chi_m}{y\pi},
\end{align}
such that they coincide with $F_0$ and $\Phi_0$ in the scaling limit.
In contrast to $F_0$ and $\Phi_0$, $F$ and $\Phi$ do not depend solely
on $y$ and $H/T_K$ but on all parameters of the Anderson model $U$,
$\varepsilon_d$, $\Gamma$ and $H$.
The range of practical validity of the scaling form Eq.~\eqref{rqscaling}
is tested below.

\begin{figure}[t!]
\begin{center}
    \includegraphics[width=\linewidth]{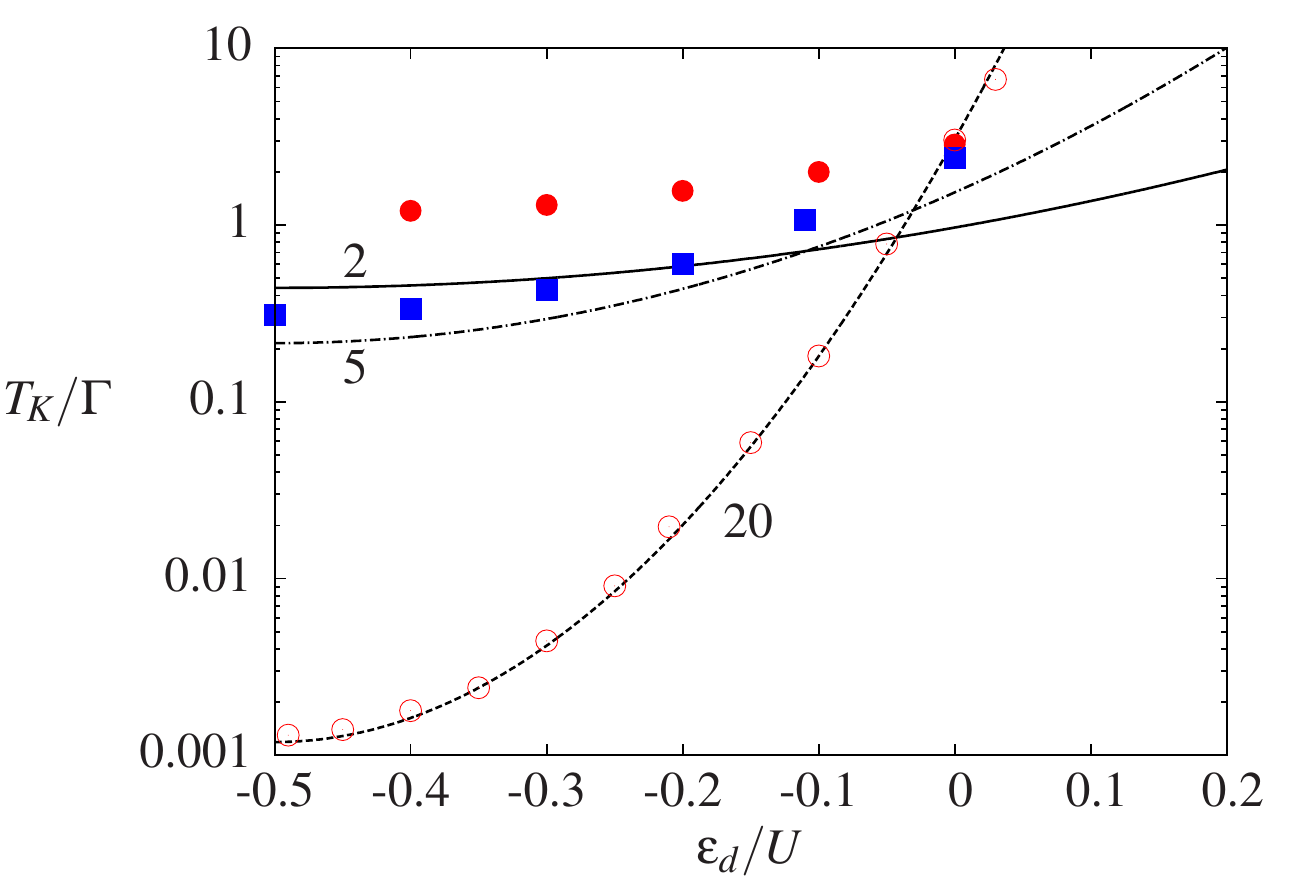}
\caption{Numerical values of $T_K$ obtained (see main text) from the BA equations for $U/\Gamma=$ 2, 5 and 20 (full circles, squares and empty circles respectively). They are compared to the analytical formula Eq.~\eqref{tk} (solid lines).}\label{fig:tk}
\end{center}
\end{figure}

\subsection{The resistance peak in the function $\Phi$}

The departure of $\Phi$ from $\Phi_0$ is studied in Fig.~\ref{fig:phi}
by plotting $\Phi$ as a function of the magnetic field $H$
for different values of $U/\Gamma$ and the asymmetry parameter $y$.
The Kondo temperature $T_K$ used to rescale the magnetic field
in Fig.~\ref{fig:phi} is obtained by numerically
matching the low field behavior
of $\Phi$ with the expected asymptotic form 
$\Phi_0 (H/T_K) \simeq H/T_K$ for $H\ll T_K$.
The result for $T_K$ is shown in Fig.~\ref{fig:tk} where it is
compared to the Kondo temperature Eq.~\eqref{tk} of the Anderson
model.

A first regime can be identified for $U/\Gamma>5$ where the universal function $\Phi_0 (H/T_K)$ is well reproduced in the Kondo region. The deviation between $\Phi$ and $\Phi_0$ becomes sizable only close to the Coulomb peaks, where $\varepsilon_d^*\simeq 0$, as seen in Fig. \ref{fig:chic}. At these charge degeneracy points, the peak in the charge relaxation resistance decreases in magnitude with $\varepsilon_d$ but does not disappear. The form of the resistance peak in the crossover from  the empty orbital to the valence fluctuation region is discussed in Sec.~\ref{sec:mixed}.

For $U/\Gamma\leq5$, Kondo physics is much less pronounced which results in a lowering of the peak in $\Phi$. The agreement between the calculated Kondo temperature using our fitting procedure and Eq. (\ref{tk}) is also degraded as shown in Fig.~\ref{fig:tk}.


\begin{figure}[b!]
\begin{center}
\subfigure[]{\includegraphics[width=\linewidth]{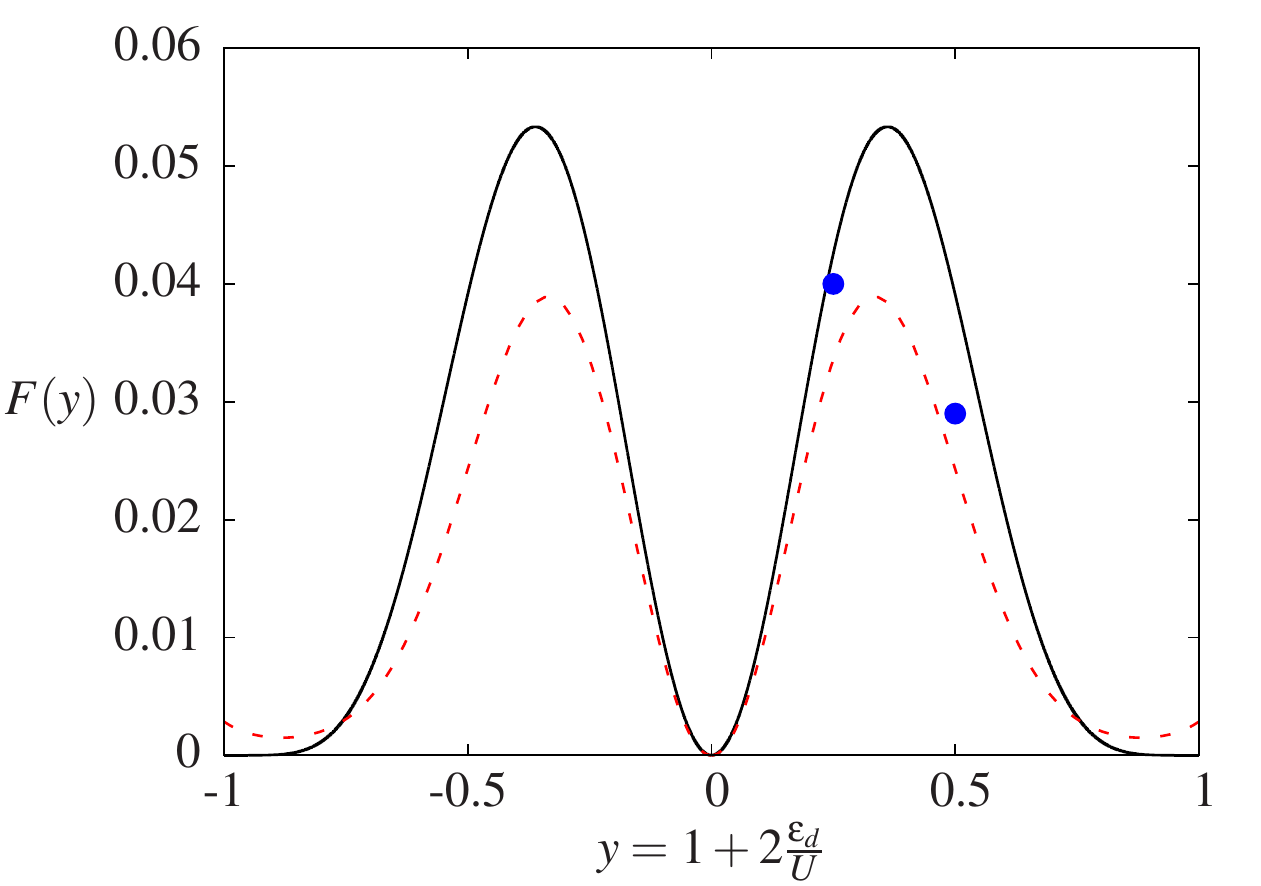}\label{Fleading}}
\subfigure[]{\includegraphics[width=\linewidth]{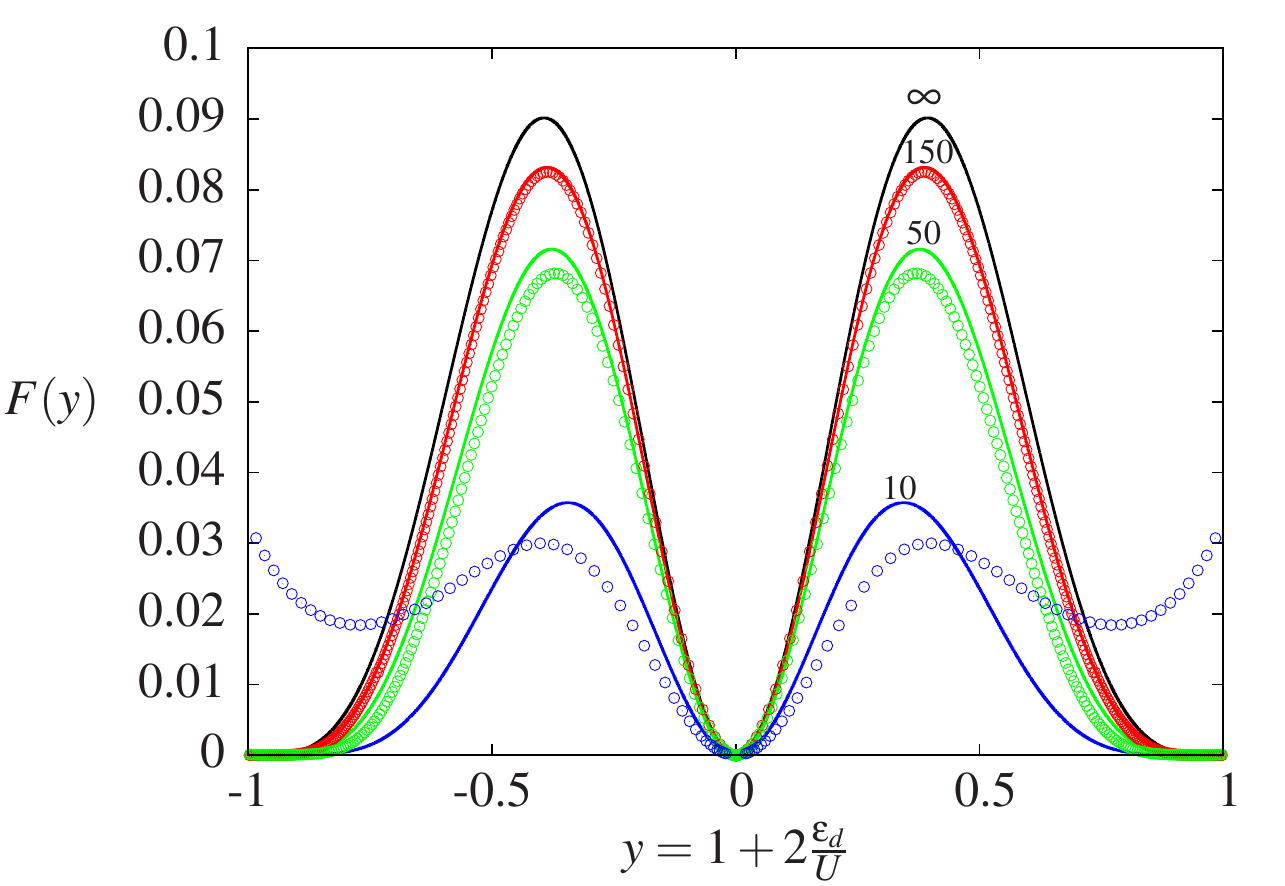}}
\caption{a) Comparison of NRG (points extracted from the results of Ref.~\cite{lee2011}), BA (dashed line) and analytical (solid line) results for the function $F$ with $U/\Gamma=20$. b) Approach to the scaling limit $F_0$ Eq. (\ref{eqFleading}) for different $U/\Gamma$. The dotted lines are obtained by BA while the solid ones correspond to the perturbative result Eq.~\eqref{chied}.}\label{fig:F}
\end{center}
\end{figure}

\subsection{The envelope function $F$ in the Kondo region}

Fig.~\ref{fig:chic} demonstrates that, as long as one remains in the Kondo region, the dependence of $\chi_c$ on the magnetic field can be safely neglected. In Fig.~\ref{fig:F}.a, our BA calculation for the $F$ function at zero magnetic field, represented by the dashed line, is in very good agreement with the NRG data extracted from Ref.~\cite{lee2011}. It remains however far from the asymptotic function $F_0 (y)$ even though $U/\Gamma=20$ in Fig.~\ref{fig:F}.a, see also Ref.~\cite{filippone2011}.

The convergence of $F$ to the asymptotic form $F_0(y)$ as a function of $U/\Gamma$ is illustrated in Fig.~\ref{fig:F}.b where it is shown to be slow. A more quantitative analytical expression for $F$ can be derived by including the next to leading order corrections to the charge susceptibility, namely \cite{filippone2012}

\begin{equation}\label{chied}
\begin{split}
\chi_c=&\frac{\Gamma}\pi\left\{\frac1{(\varepsilon_d+U)^2}+\frac1{\varepsilon_d^2}+\frac{2\Gamma}\pi\right.\left.\left[\frac1{(\varepsilon_d+U)^3}-\frac1{
\varepsilon_d^3 } \right ]
\right.\\
+&\left.\frac\Gamma\pi\left[\left(\frac1{\varepsilon_d+U}-\frac1{\varepsilon_d}\right)^3\right.\right.\\
+&\left.\left.2\left(\frac1{\varepsilon_d+U}-\frac1{\varepsilon_d}\right)\left(\frac1{
\varepsilon_d^2}-\frac1{(\varepsilon_d+U)^2}\right)\ln\frac{\varepsilon_d+U}{-\varepsilon_d}\right]\right\}.
\end{split}
\end{equation}
in Eq.~\eqref{fphi}. The result for $F$ is the function $F_0(y)$ with additional $\Gamma/U$ corrections. It is in much better agreement with the BA calculations and the NRG results from Ref.~\cite{lee2011} than $F_0$ alone, as shown in Fig.~\ref{fig:F}.a.


\section{The valence-fluctuation region}\label{sec:mixed}

The meaning of Eq.\eqref{rqscaling} is restricted to the Kondo region where a Kondo temperature can be defined.

As we already saw in Fig.~\ref{fig:phi}, the peak in the charge relaxation resistance decreases in magnitude at the edge of the Kondo region, in the mixed-valence region around $\varepsilon_d^* \simeq 0$. Below we discuss the fate of the resistance peak as $\varepsilon_d$ is further increased to explore the empty orbital region $\varepsilon_d \gg \Gamma$, and the valence-fluctuation region at higher magnetic field. As we shall see below, the resistance peak does not disappear although its magnitude does not scale with $U/\Gamma$ in this region.

The peak in the charge relaxation resistance can be derived analytically in the regime $\varepsilon_d \gg \Gamma$ by standard perturbation theory. In this regime and for arbitrary magnetic field, the two states of the isolated dot forming the low energy sector are $\ket{n=0}$ and $\ket{n=1,\uparrow}$ as shown in Fig.~\ref{fig:levels}. The absence at low energy of the spin down component implies that the ground state does not exhibit strong correlation and can be described analytically using perturbation theory. The unperturbed Hamiltonian is obtained by setting the tunneling involving spin down electrons 
\begin{equation}\label{tunneling}
t \sum_k \left( c^\dagger_{k\downarrow} d_\downarrow + d_\downarrow^\dagger
c_{k\downarrow} \right)
\end{equation}
to zero. In that case, the number of spin down electrons on the dot is a constant of motion and the Hamiltonian can be diagonalized separately for $n_\downarrow = 0$ (low energy) and $n_\downarrow = 1$ (high energy). It gives an exactly solvable resonant level model 
\begin{equation}\label{resonant}
H'=\sum_{k}\varepsilon_kc^\dagger_{k\uparrow} c_{k\uparrow}+\left( \varepsilon_{d\uparrow} + U n_\downarrow \right) n_\uparrow + t \sum_k\left(c^\dagger_{k\uparrow}d_{\uparrow}+d^\dagger_\uparrow c_{k\uparrow}\right),
\end{equation}
for which the charge relaxation resistance is $h/2e^2$.

Let us call $| \psi_0 \rangle$ the unperturbed ground state, with $n_\downarrow = 0$, characterized by the spin up electron occupancy on the dot
\begin{equation}
\langle n_\uparrow \rangle_0 = \langle \psi_0 | n_\uparrow | \psi_0 \rangle
= \frac 1 2 - \frac{1}{\pi} \, {\rm arctan} \left( \frac{\varepsilon_d-\frac H2}
{\Gamma} \right).
\end{equation}
The perturbation due to the tunneling term Eq.~\eqref{tunneling}
gives the first order correction to the wave function
\begin{equation}\label{ground}
\begin{split}
\ket{\psi_1}=&t\sum_k\left(\frac1{\varepsilon_k-U-\varepsilon_d-\frac H2}d^\dagger_\downarrow c_{k\downarrow}n_\uparrow+\right.
\\&+\left.\frac1{\varepsilon_k-\varepsilon_d-\frac H2}d^\dagger_\downarrow c_{k\downarrow}(1-n_\uparrow)\right)\ket{\psi_0}.
\end{split}
\end{equation}
The projectors $n_\uparrow$ and $(1-n_\uparrow)$ are necessary to determine the part of $\ket{\psi_0}$ with a spin up electron on the dot and the part with no electron. This implies the presence or not of the interaction energy $U$ in the denominator of Eq.~\eqref{ground}.
\begin{figure}
\includegraphics[width=\linewidth]{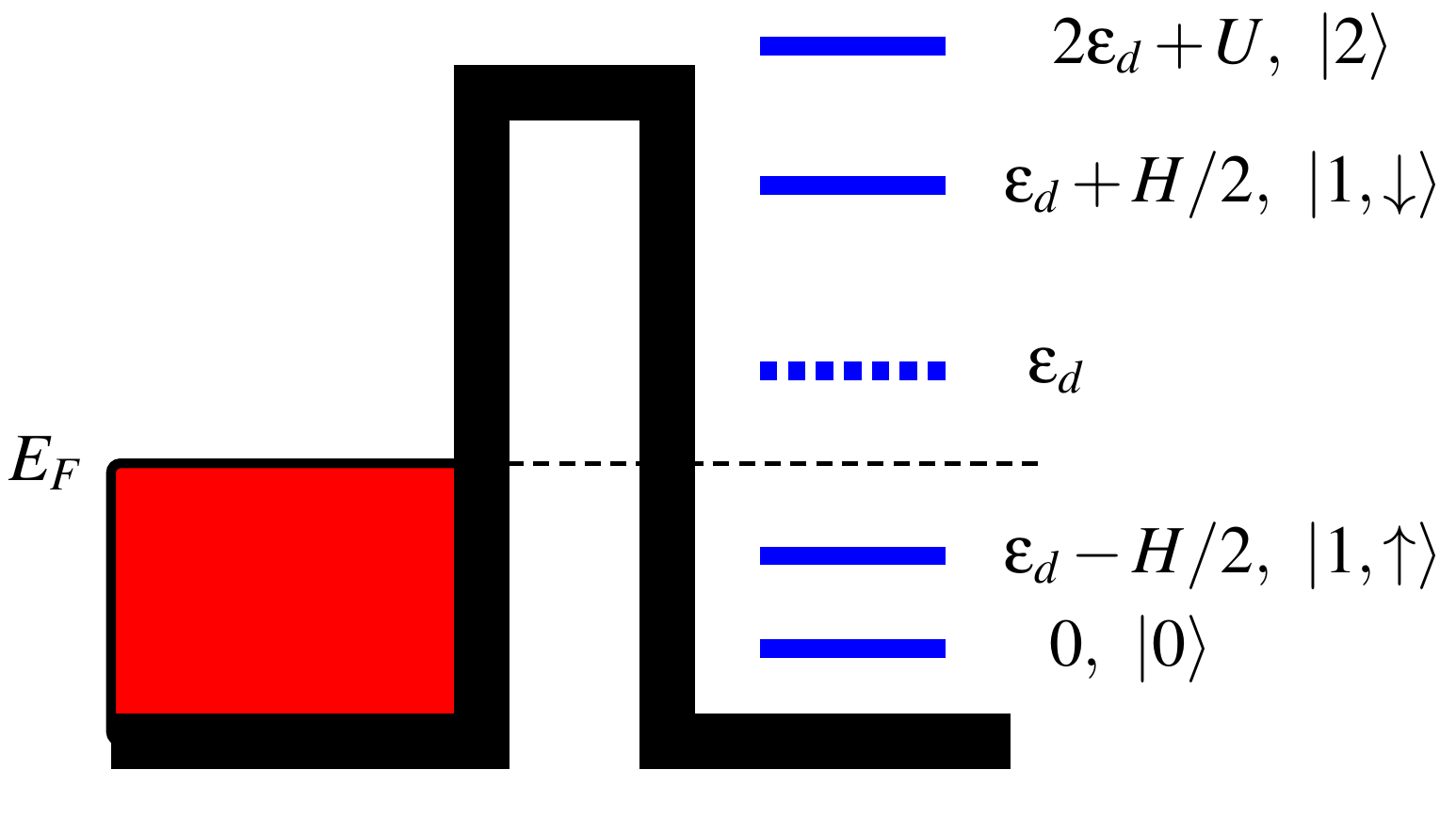}
\caption{Spectrum of the dot isolated from the lead on the left. For a positive $\varepsilon_d$ in the presence of a magnetic field, only the states $\ket0$ and $\ket{1,\uparrow}$ compete in the low energy sector.}\label{fig:levels}
\end{figure}
 The values of the spin $\sigma$ populations for the corrected ground state $\ket{\psi_0}+\ket{\psi_1}$ are
\begin{equation}
\begin{split}
\av{n_\uparrow}&=\av{n_\uparrow}^0-\frac\Gamma\pi \frac{U\av{n_\uparrow}^0(1-\av{n_\uparrow}^0)}{(U+\varepsilon_d+\frac H2)(\varepsilon_d+\frac H2)},\\
\av{n_\downarrow}&=\frac\Gamma\pi\left(\frac{1-\av{n_\uparrow}^0}{\varepsilon_d+\frac H2}+\frac{\av{n_\uparrow}^0}{\varepsilon_d+U+\frac H2}\right),
\end{split}
\end{equation}
corresponding to the static susceptibilities
\begin{align}\label{pertsusc}
\begin{split}
\chi_\uparrow=&\chi_\uparrow^0-\frac\Gamma\pi\frac{\chi_\uparrow^0(1-2\av{n_\uparrow}^0)U}{(\varepsilon_d+U+H/2)(\varepsilon_d+H/2)}-\\
&\frac\Gamma\pi\frac{\av{n_\uparrow}^0(1-\av{n_\uparrow}^0)[U^2+2U(\varepsilon_d+H/2)]}{(\varepsilon_d+U+H/2)^2(\varepsilon_d+H/2)^2},\\
\chi_\downarrow=&\frac \Gamma\pi\left\{\frac{1-\av{n_\uparrow}^0}{(\varepsilon_d+\frac H2)^2}+\frac {\av{n_\uparrow}^0}{(\varepsilon_d+\frac H2+U)^2}+\right.\\
&\left.~~~~\chi_\uparrow^0\left(\frac {1}{\varepsilon_d+\frac H2+U}-\frac{1}{\varepsilon_d+\frac H2}\right)\right\}.
\end{split}
\end{align}
We have introduced
\begin{equation}\label{lorentzian}
\chi_\uparrow^0=\frac\Gamma\pi\frac1{\left(\varepsilon_d-H/2\right)^2+\Gamma^2},
\end{equation}
the spin up susceptibility in the absence of the spin down component. 

The static susceptibilities of Eq.~\eqref{pertsusc} are combined to give $\chi_c=\chi_\uparrow+\chi_\downarrow$ and $\chi_m=\chi_\uparrow-\chi_\downarrow$. Substituted in Eq.~\eqref{rqintro}, they give an analytical expression for the charge relaxation resistance $R_q$ which still exhibits a peak as a function of the magnetic field, as shown in Fig.~\ref{fig:pert}. Fig.~\ref{fig:pert} also compares the analytical expression for $R_q$ with the BA calculations and shows an excellent agreement already for  $\varepsilon_d/\Gamma=6$. The peak height occurs around $h/2 e^2$ and for $H \simeq 2 \varepsilon_d$. At this point, the spin up charge fluctuations are maximum, see Eq.~\eqref{lorentzian}, because the states $\ket{n=0}$ and $\ket{n=1,\uparrow}$ are degenerate for the isolated dot when $H =2 \varepsilon_d$, and the spin down fluctuations remain small. Hence, the resistance is around $h/2 e^2$ as in the single-channel spinless case. The position of the maximum of the resistance can be found perturbatively from the analytical solution
Eq.~\eqref{pertsusc}
\begin{align}
\frac H{2 \varepsilon_d }= &1-\frac{\Gamma}{\pi \varepsilon_d}\frac{U(4 \varepsilon_d+U)}{(2 \varepsilon_d+U)^2},\\
R_q =&\frac{h}{2e^2}\left(1+\frac{\Gamma}{\pi \varepsilon_d}
\frac{U}{2 \varepsilon_d+U}\right).
\end{align}
\begin{figure}[t!]
\begin{center}
\includegraphics[width=\linewidth]{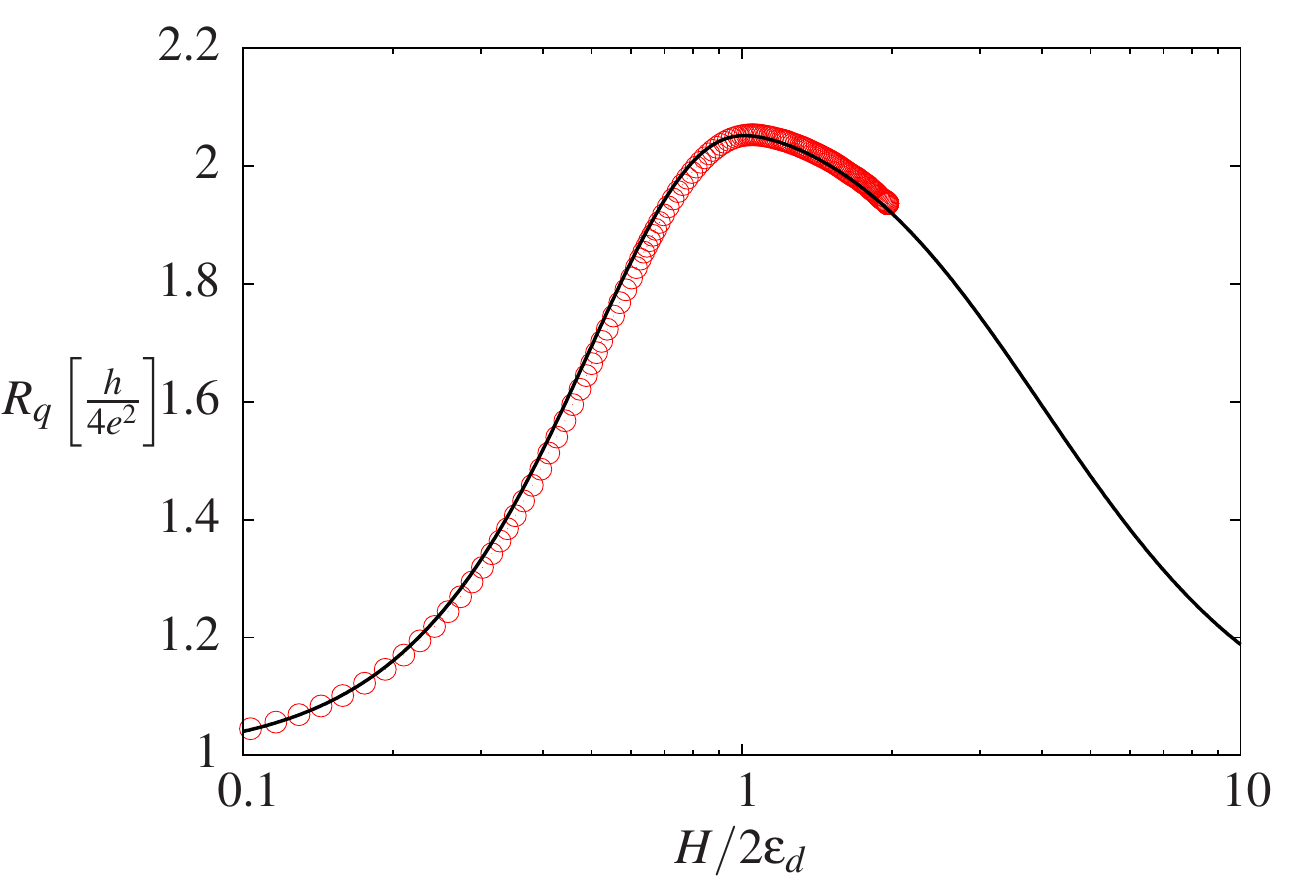}
\caption{Comparison between $R_q$ obtained from the analytical results Eqs.~\eqref{pertsusc} (solid line) and the numerical solution of the BA equations (circles) for $U/\Gamma=20$ and $\varepsilon_d/\Gamma=6$.}\label{fig:pert}
\end{center}
\end{figure}

The analytical expression obtained for $R_q$ from  Eq.~\eqref{pertsusc} can be further simplified in the limit $\varepsilon_d \gg \Gamma$. For $x=H/2\varepsilon_d < 1$, the universal form
\begin{equation}\label{scaling2}
R_q =\frac h{4e^2}\left[1+\frac{4x^2}{(x^2+1)^2}\right]
\end{equation}
is obtained. This result is independent of $U$ because the unperturbed ground state is $\ket 0$ when $\Gamma$ is sent to zero. The doubly occupied state is therefore reached only to second order in perturbation theory and can be neglected to leading order. For $x>1$, the unperturbed ground state is  $\ket{1,\uparrow}$ for a vanishing $\Gamma$ and the form of $R_q$ depends on the ratio $\varepsilon_d/U$. For $\varepsilon_d \gg U$, we recover essentially a non-interacting resonant level model and the resistance is also given by the universal form Eq.~\eqref{scaling2} for $x>1$. For $\varepsilon_d \ll U$ however, the charge relaxation resistance $R_q$ is frozen to $h/2 e^2$ for all $x>1$. Both these universal limits are shown in Fig. \ref{fig:scaling}. The reason is that the doubly occupied state is forbidden for infinite $U$ such that the spin down states cannot be reached within first order perturbation theory. Hence we only have spin up charge fluctuations, $\chi_{\downarrow} \to 0$, and we recover the universal result of the spinless case $R_q = h/2 e^2$. 


\section{The SU(4) Kondo case}\label{sec:carbon}

We extend our discussion to the more exotic case of a SU(4) Kondo effect~\cite{borda2003,*lehur2003,*zarand2003kondo,*lopez2005}. This situation is relevant for certain quantum dots with an additional orbital degree of freedom that is conserved during lead-dot  tunneling processes~\cite{minot2004determination}. For example, ultra-clean carbon nanotubes have a natural orbital degeneracy that arises from the clockwise and anti-clockwise motions of electrons around the tube. We label here the orbital index by $l=+,~-$. The model has now four transport channels in correspondence with the four available single-electron states in the dot: $\ket{+,\uparrow},~\ket{+,\downarrow},~\ket{-,\uparrow}\mbox{ and }\ket{-,\downarrow}$. We label these four states by a quantum number $\tau=1,\ldots,4$ respectively and use the same index for the  conduction electrons in the lead. The Hamiltonian takes the form of a SU(4) Anderson model~\cite{choi2005}:
\begin{equation}
\begin{split}
H=&\sum_{k\tau}\varepsilon_k c^\dagger_{k\tau}c_{k\tau}+t\sum_{k\tau}\left(c^\dagger_{k\tau}d_{\tau}+d^\dagger_\tau c_{k\tau}\right)\\
&+\varepsilon_d\sum_{\tau}n_\tau+U\sum_{\tau<\tau'}n_\tau n_{\tau'},
\end{split}
\end{equation}
where the meaning of the operators and notations are the same as in Eq.~\eqref{am}. For temperatures much below the interaction energy $U$ and $\Gamma\ll U$, the charge on the dot is frozen to 1, 2 or 3 depending on the gate voltage $\varepsilon_d$. Performing a Schrieffer-Wolff transformation~\cite{schrieffer1966}, one finds
\begin{equation}\label{sw}
\begin{split}
H_{SW}=&\sum_{k\tau}\varepsilon_kc^\dagger_{k\tau}c_{k\tau}+W_q\sum_{kk'\tau}c^\dagger_{k\tau}c_{k'\tau}\\
&+\frac{J_q}2\sum_{kk'\tau\tau'}c^\dagger_{k\tau}c_{k'\tau'}\left(d^\dagger_{\tau'}d_\tau-\frac qN\delta_{\tau\tau'}\right),
\end{split}
\end{equation}
where $q$ denotes the dot occupancy in the low energy sector and $N=4$. The generalization to any $N$ and $q$ is straightforward. The values of the potential scattering and the Kondo coupling constants $W_q$ and $J_q$ are given by
\begin{align}
\label{j}J_q&=-2t^2\left(\frac1{\varepsilon_d+(q-1)U}-\frac1{\varepsilon_d+qU}\right),\\
\label{w}W_{q}&=-\frac{t^2}N\left(\frac{q}{\varepsilon_d+(q-1)U}+\frac{N-q}{\varepsilon_d+qU}\right).
\end{align}
The potential scattering term vanishes for $\varepsilon_{dW0}=(1-q-q/N)U$. An exact mapping  to the SU(N) Kondo model~\cite{mora2009} is then obtained
\begin{equation}\label{su4}
H_{SU(N)}=\sum_{k\tau}\varepsilon_kc^\dagger_{k\tau}c_{k\tau}+J'_q\mathbf{S}\cdot\mathbf{T},
\end{equation}
where $J'_q=\frac{2t^2}U\frac{N^2}{q(N-q)}$. We switched to the basis of generators of SU(N) \cite{parcollet1998,jerez1998,mora2009}, such that an anti-ferromagnetic coupling between the spin $\mathbf S=\sum_{\tau \tau'}d^\dagger_\tau \frac{\mathbf \lambda_{\tau \tau'}}2d_{\tau'}$ of the impurity and $\mathbf T=\sum_{kk'\tau \tau'}c^\dagger_{k\tau} \frac{\mathbf \lambda_{\tau \tau'}}2c_{k'\tau'}$ of the lead is made explicit.  $\lambda$ is the vector composed of the $N^2-1$ matrices which compose the $N\times N$ fundamental representation of the SU(N) group. Their explicit expression in the SU(4) case can be found in \cite{greiner}.

As mentioned in the Introduction, the low energy fixed point of the Hamiltonian Eq.~\eqref{sw} is a Fermi liquid and the Fermi liquid approach~\cite{filippone2011,filippone2012} introduced in Sec.~\ref{sec:fermi} is also applicable to this model. Defining $\chi_\tau=-\partial\av{n_\tau}/\partial\varepsilon_d$ as the $\tau$-dependent static susceptibilities, the charge relaxation resistance is found to be
\begin{equation}\label{rqsum}
R_q=\frac{h}{2e^2}\frac{\sum_{\tau}\chi_\tau^2}{(\sum_\tau\chi_\tau)^2}.
\end{equation}
The emergence of logarithmic singularities prevents the study of the $\chi_\tau$ susceptibilities by perturbative methods below the SU(4) Kondo temperature \cite{choi2005}
\begin{equation}\label{tksu4}
T_K^{q}=\mathcal D e^{-1/(2\nu_0J_q)},
\end{equation}
where $\mathcal D \simeq U,\varepsilon_d$ is the effective high-energy cut-off of the model whose precise form is not needed here.

Following the line of reasoning developed in Ref.~\cite{filippone2011}, one can derive the  behavior of $R_q$ in the presence of a magnetic field. We first switch to a more convenient basis that separates the charge, spin and orbital degrees of freedom, namely
\begin{equation}\label{change}
\left(
\begin{array}{c}
\chi_c\\
\chi_m\\
\chi_v\\
\chi_{mv}
\end{array}\right)=
\left(
\begin{array}{cccc}
1&1&1&1\\
1&-1&1&-1\\
1&1&-1&-1\\
1&-1&-1&1
\end{array}
\right)\left(
\begin{array}{c}
\chi_1\\\chi_2\\\chi_3\\\chi_4
\end{array}\right).
\end{equation}
In addition to the total charge susceptibility $\chi_c$, we have introduced the
charge magneto-susceptibility $\chi_m$, as in the SU(2) case, and its orbital counterpart, $\chi_v$, which measures the sensitivity of the orbital magnetization to a change in gate voltage. $\chi_{mv}$ is obtained from the difference between the spin magnetizations of the two orbital states.

Substituting the new susceptibilities in Eq.~\eqref{rqsum}, 
the charge relaxation resistance is found to be 
\begin{equation}\label{rqgen}
R_q=\frac{h}{8e^2}\left(1+\frac{\chi_m^2+\chi_v^2+\chi_{vm}^2}{\chi_c^2}\right),
\end{equation}  
the analog of Eq.~\eqref{rqintro} in the SU(4) case.
At zero magnetic field, the spin and orbital degeneracies are not broken such that $\chi_m = \chi_v = \chi_{vm}=0$ and a universal resistance $R_q=h/8e^2$ is obtained. At finite magnetic field, only the spin degeneracy is broken and $\chi_v = \chi_{vm}=0$.

In the limit $U \gg \Gamma$ and for magnetic fields of the order of the Kondo temperature Eq.~\eqref{tksu4}, the magnetic field dependence of the charge susceptibility $\chi_c$ can be neglected. Assuming that the results of Cragg and Llyod~\cite{cragg1978} are also valid in the SU(4) case, such that the leading potential scattering term in Eq.~\eqref{sw} is unaltered along the Kondo crossover, the Friedel sum rule leads to 
\begin{equation}\label{chic}
\chi_c=4\nu_0 \partial_{\varepsilon_d} W_m = \frac\Gamma \pi
\left[ \frac{q}{(\varepsilon_d + (q-1) U)^2} + 
\frac{4-q}{(\varepsilon_d + q U)^2} \right]
\end{equation}
in the sector with $q$ charges on the dot.
\begin{figure}
\includegraphics[width=\linewidth]{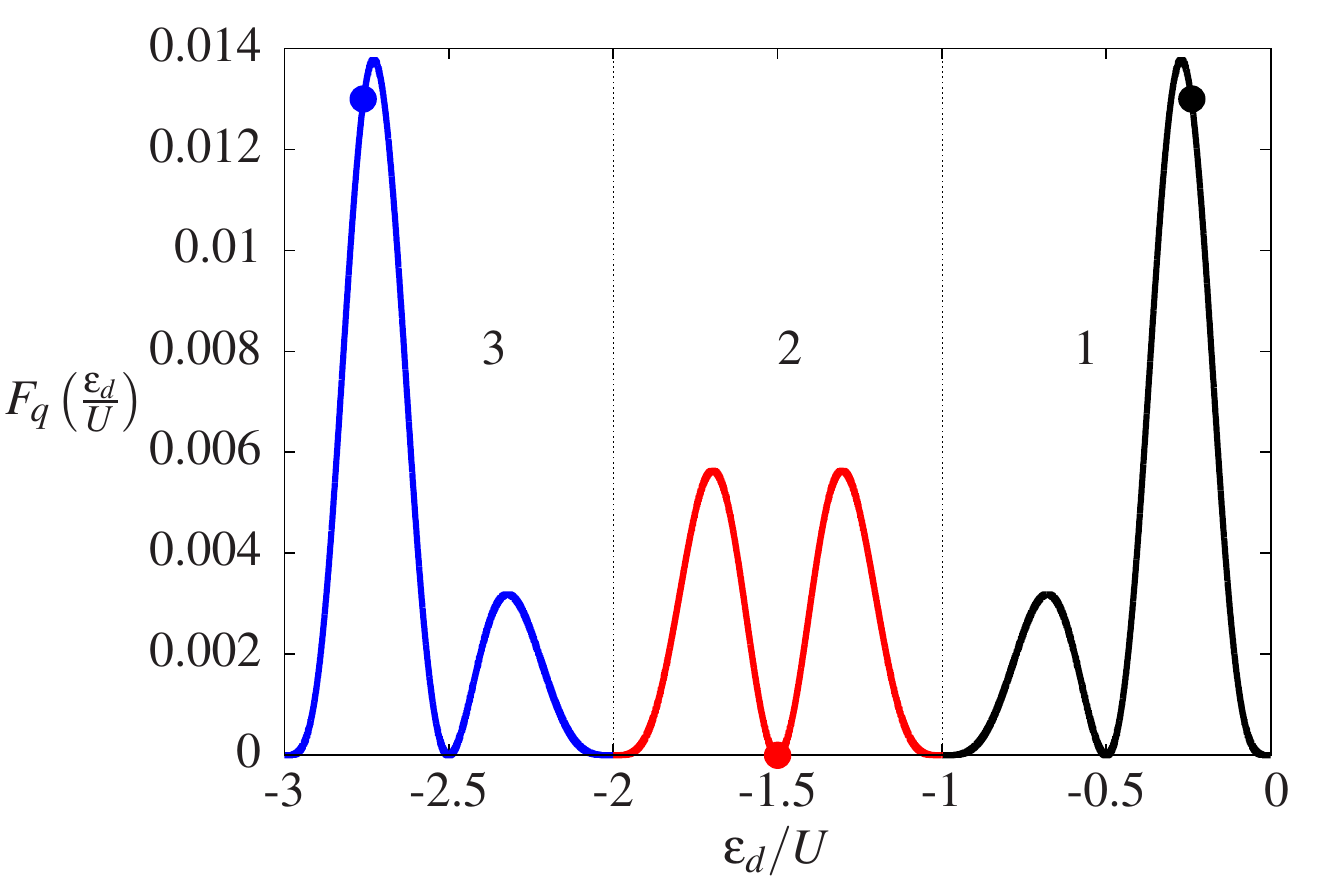}
\caption{Envelope functions $F_q$ in the sectors where the charge is frozen to 1, 2 or 3, that is when $\varepsilon_d/U~\in$ [-1,0], [-2,-1] or [-3,-2] respectively.
The function becomes zero in the middle of the Coulomb valleys, while the circles correspond to the values of $\varepsilon_d/U$ for which the potential scattering couplings $W_q$ in Eq. (\ref{w}) become zero.}\label{fsu4}
\end{figure}

As in the SU(2) case, the form of the charge magneto-susceptibility $\chi_m$ can be derived from scaling arguments. In the Kondo limit, the magnetization of the dot $m=\sum_{l,\sigma}\sigma\av{n_{l\sigma}}/2$ has been derived from the Bethe ansatz solution of the SU(N) Kondo Hamiltonian Eq. \eqref{su4} \cite{bazhanov2003,coqblin1969}. It is a smooth and monotonous universal function  $f_q(H/T^{q}_K)$ that starts at zero at vanishing magnetic field and saturates at $1/2$ (resp. $1$) for large magnetic fields, when $q=1,3$ (resp. $q=2$). Differentiating the magnetization with respect to $\varepsilon_d$, one obtains 
\begin{equation}\label{chicq}
\chi_m=2 \partial_{\varepsilon_d}\ln T_K^{q} \, \,\Phi_q (H/T_K^q)
\end{equation}
where we defined the universal functions $\Phi_q (x) = x f_q'(x)$. From the general form of the functions $f_q(x)$, we expect that the functions $\Phi_q(x)$ have a similar peaked shape as the function $\Phi(x)$ of the SU(2) case. Using the expression Eq.~\eqref{tksu4} of the Kondo temperature, we obtain to leading order in $\Gamma/U$
\begin{equation}\label{chim}
\chi_m= \frac{\pi}{2 \Gamma} \, \frac{2 \varepsilon_d + (2 q -1) U}{U} \,
\Phi_q \left(\frac H{T_K^{q}}\right), 
\end{equation}
where the prefactor $2 \varepsilon_d + (2 q -1) U$ 
essentially comes from the derivative of the Kondo temperature Eq.~\eqref{tksu4}.
Combining Eqs.~\eqref{chic} and~\eqref{chim} into Eq.~\eqref{rqgen}, we find a scaling law in the Kondo limit 
\begin{equation}\label{rqsu4}
R_q=\frac h{8e^2}\left[1+\left(\frac U\Gamma \right)^4F_q\left(y_q\right)\Phi^2_q\left(\frac{H}{T_K^q}\right)\right]
\end{equation}
similar to the SU(2) case. Thus a giant peak in the charge relaxation resistance, proportional to $(U/\Gamma)^4$, also emerges for a SU(4) symmetry. The envelope functions 
\begin{equation} 
F_q (y_q)=  \left( \frac{\pi^2}{32} \right)^2
\,  \frac{ y_q^2 \left(y_q^2-1\right)^4}{[1+y_q^2+y_q (q-2)]^2}, 
\end{equation}
depend on the charge $q$ and on the variable $y_q = 2 \varepsilon_d/U + 2 q -1$. $y_q$ is defined such that $y_q=\pm 1$ at the Coulomb peaks and $y_q=0$ in the middles of the Coulomb valleys. The envelope functions corresponding to the three charge sectors $q=1,2,3$ are represented on the same plot Fig. \ref{fsu4} as a function of $\varepsilon_d/U$. 

Interestingly, the function $F_2$ coincides with the SU(2) function $F$ up to the multiplicative factor $16$. Instead, in the sectors $q=1$ and 3, the envelope function is asymmetric, which gives an experimental signature distinguishing SU(2) and SU(4) symmetries. We also notice that the values of $\varepsilon_d/U$, for which the envelope functions $F_{1,3}$ vanish, do not coincide with the locations of zero potential scattering, {\it i.e.} $W_q=0$ in Eq.~\eqref{w}, represented by circles in Fig.~\ref{fsu4}. We expect that the approach to the Kondo scaling behavior is faster at those latter points since they are free of potential scattering and exhibit only Kondo coupling. In addition, the envelope is close to its maximum at these points, in contrast with $q=2$ and the SU(2) case where the envelope vanishes as imposed by particle-hole symmetry.
\vspace{3mm}

As a final remark before concluding, we stress that the discussion above can be generalized to the case of an extended SU(N) symmetry. The Fermi liquid picture still holds in that case\cite{mora2009a}, and Eq.~\eqref{rqsum}, with $\tau=1,\ldots,N$, predicts the universal result
\begin{equation}\label{symmetry}
R_q = \frac{h}{2 N e^2}
\end{equation}
if all channels are symmetric. Indeed, in the symmetric case, $\chi_\tau = \chi_c/N$. $\chi_c=\sum_\tau\chi_\tau$ is the total charge susceptibility and appears in the denominator of Eq. \eqref{rqsum}.
In the channel-asymmetric case, the transformation Eq.~\eqref{change} extends in the following way 
\begin{equation}
\left(\begin{array}{c}
\chi_c\\
\chi'_1\\
\vdots\\
\chi'_{N-1}
\end{array}\right)=
\left(\begin{array}{ccccc}
1&1&\ldots&\ldots&1\\
&&v_1&&\\
&&\vdots&&\\
&&v_{N-1}&&\\
\end{array}\right)
\left(\begin{array}{c}
\chi_1\\\chi_2\\\vdots\\\chi_N
\end{array}\right)\,.
\end{equation}
The first row vector $(1,1,\ldots,1)$ of the transformation matrix gives $\chi_c$. The remaining vector $v_i$ depend on the specific problem, they are however orthogonal to the first row vector and normalized to $N$. The resulting expression for the charge relaxation resistance reads
\begin{equation}
R_q=\frac h{2Ne^2}\left(1+\frac{\sum_{i=1}^{N-1}\chi_i'^2}{\chi_c^2}\right)\,.
\end{equation}
generalizing Eq.~\eqref{rqgen}. A coupling to one of the vector $v_i$ (such as a magnetic field or an orbital energy term) breaks the channel symmetry and should be responsible for a similar peak in the charge relaxation resistance on energy scales on the order of the SU(N) Kondo temperature.


\section{Conclusions}  

In this paper, we performed a thorough study of the quantum capacitance and the charge relaxation resistance for the Anderson model. We applied a Fermi liquid approach, where the low energy effective model is derived consistently with the Friedel sum rule, that allowed us to express the charge relaxation resistance in terms of static susceptibilities. The susceptibilities are computed from the Bethe ansatz equations describing the ground state of the Anderson model. The accuracy of our approach was tested by comparing our results to NRG calculations~\cite{lee2011} or perturbative calculations both in the Kondo and in the strongly asymmetric regimes. The analytical predictions given in Ref.~\cite{filippone2011} for the peak in the charge relaxation resistance are shown to apply in the whole Kondo region for $U/\Gamma>5$. The persistence of this peak was demonstrated in the valence-fluctuation region, both numerically and from a direct perturbative calculation. Moreover, we showed how the Fermi liquid approach can be extended to the SU(4) symmetric case where a similar peak emerges in the charge relaxation resistance.

Overall, this work constitutes a specific and detailed example of how the effective Fermi liquid theory can be used to derive the low frequency dynamics of quantum impurity systems. This does not include, of course, systems and regimes in which non-Fermi liquid physics~\cite{mora2012low,dutt2013} dominates such as impurity models with overscreening. We also mention the possibility to apply Eq. \eqref{rqintro} to the case of a multi-level quantum dot \cite{yeyati1999} with spin 1/2 electrons in the lead. The Friedel sum rule applies in these systems \cite{rontani2006} and non-monotonous behaviors are expected to emerge in the charge relaxation resistance
$R_q$ whenever the magnetization of the quantum dot varies substancially with $\varepsilon_d$, leading to $\chi_m\neq0$. This includes notably the breaking of the Kondo singlet in the presence of a magnetic field also in the multi-level case.

Further extensions of this work could include the study of non-zero temperatures and higher frequencies~\cite{lee2011,crepieux2012} where inelastic processes play an increasing role. Quite generally, the main effect of finite temperature is to destroy quantum coherence of electrons in the dot leading to a convergence of the charge relaxation resistance with the DC resistance~\cite{nigg2008quantum,rodionov2009}. The analysis of this paper relies essentially on the generalized Korringa-Shiba relation Eq.~\eqref{korringa}, which is strictly valid only at zero temperature. Finite temperature effects could be addressed quantitatively by including Nozi\`eres' Fermi liquid corrections to the fixed point \cite{nozieres1974fermi,mora2009a}. This would modify Eqs. \eqref{korringa} and \eqref{rqintro}. Qualitatively, the peak in the charge relaxation resistance should survive for temperatures below the Kondo temperature. Above the Kondo temperature, the Kondo singlet is completely broken and the form of $R_q$ with the magnetic field remains an open question left for further study.

We acknowledge T. Kontos for useful discussions and thank the authors of Ref.~\cite{lee2011} for providing us their NRG data. KLH acknowledges support from DOE under the grant DE-FG02-08ER46541.


\appendix
\begin{widetext}
\section{Bethe ansatz equations for the ground state of the Anderson model}\label{app:BA}
A striking feature of one dimensional quantum systems \cite{giamarchi2004quantum} is the possibility to have a separation between charge and spin degrees of freedom for electrons at low temperature. In the case of the Anderson model, spin and charge are carried by different excitations called $spinons$ and $holons$ respectively. Their densities of states are denoted $\rho$ and $\sigma$. They satisfy the following Bethe ansatz integral equations \cite{wiegmann1983,tsvelick1983,kawakami1982ground} (we follow the notations of Ref.~\cite{kawakami1982ground}):
\begin{align}
\label{betherho} \rho(k)+g'(k)\int_{-\infty}^Bdp\rho(p)R[g(k)-g(p)]+g'(k)\int_{-\infty}^Qd\lambda\sigma(\lambda)s[g(k)-\lambda] & = \mathcal S^\rho(k),\\
\label{bethesigma}  \sigma(\lambda)-\int_{-\infty}^Qd\lambda'\sigma(\lambda')R(\lambda-\lambda')+\int_{-\infty}^Bdk\rho(k)s[\lambda-g(k)] & =  \mathcal S^\sigma(\lambda),
\end{align}
with the source terms given by
\begin{align}
\mathcal S^\rho(k)&=\frac1{2\pi}\left\{1+g'(k)\int_{-\infty}^\infty dpR[g(k)-g(p)]\right\}+\frac1L\left\{\Delta(k)+g'(k)\int_{-\infty}^\infty dp\Delta(p)R[g(k)-g(p)]\right\},\\
\mathcal S^\sigma(\lambda)&=\int_{-\infty}^\infty dk s(\lambda-g(k))\left[\frac1{2\pi}+\frac{\Delta(k)}L\right].
\end{align}
We have introduced the functions
\begin{align}
R(x)&=\frac1{2\pi}\int_{-\infty}^\infty d\omega\frac{e^{-i\omega x}}{1+e^{|\omega|}}, & s(x)&=\frac1{2\cosh(\pi x)},\\
g(k)&=\frac{k-\varepsilon_d-U/2}{2U\Gamma}, & \Delta(k)&=\frac\Gamma\pi\frac1{(k-\varepsilon_d)^2+\Gamma^2}.
\end{align}
$L$ is the size of the system and the holon and spinon densities can be split in a conduction and impurity (dot) part
\begin{align}
\rho(k)&=\rho_c(k)+\frac{\rho_i(k)}L,  &  \sigma(\lambda)=\sigma_c(\lambda)+\frac{\sigma_i(\lambda)}L.
\end{align}
The linearity of Eqs. (\ref{betherho}) and (\ref{bethesigma}) implies that the conduction and impurity terms decouple. The former fixes the macroscopic properties of the system, {\it i.e.} the global magnetic field $H$ and the position of the valence level $\varepsilon_d$,
\begin{align}
\frac H{2\pi}&=\int^{B}_{-\infty}dk\rho_c(k), & \frac1\pi\left(\varepsilon_d+\frac U2\right)&=\int_{-\infty}^Q d\lambda \sigma_c(\lambda),
\end{align} 
while the latter gives the occupancy $\av{n}$ and the magnetization $\av{m}$ of the dot, namely
\begin{align}
\av m&=\frac12\int_{-\infty}^B dk\rho_i(k), & \av n&=1-\int_{-\infty}^Q d\lambda \sigma_i(\lambda).
\end{align} 
These equations hold exclusively for $\varepsilon_d\geq-U/2$ and $H\geq0$, while the results for $\varepsilon_d<-U/2$ are obtained by particle-hole symmetry. 

The zero magnetic field case $H=0$ and the particle-hole symmetric point $\varepsilon_d=-U/2$ are obtained by setting $B$ and $Q$ respectively to $-\infty$. In these cases, the BA equations for $\rho$ and $\sigma$ decouple and an analytical solution can be constructed on the basis of the Wiener-Hopf method~\cite{tsvelick1983}. 
\end{widetext}

\bibliographystyle{apsrev4-1}
\bibliography{biblio}

\end{document}